\documentclass[fleqn,usenatbib]{mnras}
\usepackage{newtxtext,newtxmath}
\usepackage[T1]{fontenc}
\usepackage{ae,aecompl}
\usepackage{graphicx}	
\usepackage{amsmath}	
\usepackage{amssymb}	
\usepackage{xspace}
\usepackage{booktabs}
\usepackage[flushleft]{threeparttable}
\usepackage{multirow}
\graphicspath{ {Images/} }

\newcommand{\teff}{T$_{\mathrm{eff}}$}
\newcommand{\logg}{log~$g$}
\newcommand{\feh}{[Fe/H]}
\newcommand{\kms}{km~s$^{-1}$}

\newcommand{\vmicro}{$\xi$}

\newcommand{\sname}{Psc--Eri}
\newcommand{\numstar}{42}
\newcommand{\numel}{22}

\newcommand{\Msun}{M\textsubscript{\(\odot\)}}

\newcommand{\project}[1]{\textsl{#1}}
\newcommand{\gaia}{\project{Gaia}}

\newcommand{\ith}{\ensuremath{^{\rm th}}}
\newcommand{\caiihk}{Ca~{\sc ii} H \& K}

\title[Chemical Nature of the Psc--Eri Stream]{The Chemical Nature of the Young 120-Myr-old Nearby Pisces--Eridanus Stellar Stream Flowing through the Galactic Disk}

 \author[Hawkins et al. 2020]{Keith~Hawkins$^{1}$\thanks{E-mail: keithhawkins@utexas.edu}, Madeline~Lucey$^{1}$, and Jason Curtis$^{2}$ \\
$^{1}$Department of Astronomy, The University of Texas at Austin, 2515 Speedway Boulevard, Austin, TX 78712, USA\\
$^{2}$Department of Astrophysics, American Museum of Natural History, Central Park West, New York, NY, USA
 }
\date{Accepted 2020 June 8. Received 2020 May 23; in original form 2020 March 10}

\pubyear{2020}

\begin{document}
\label{firstpage}
\pagerange{\pageref{firstpage}--\pageref{lastpage}}
\maketitle

\begin{abstract}
Recently, a new cylindrical-shaped stream of stars up to 700~pc long was discovered hiding in the Galactic disk using kinematic data enabled by the \gaia\ mission. This stream of stars, 
dubbed Pisces--Eridanus (Psc--Eri), was initially thought to be as old as 1~Gyr, yet its stars shared a rotation period distribution consistent with a population that was 120-Myr-old. Here, we explore the detailed chemical nature of this stellar stream. We carried out high-resolution spectroscopic follow-up of \numstar\ \sname\ stars using McDonald Observatory, and combined these data with information for 40 members observed with the low-resolution LAMOST spectroscopic survey. Together, these data enabled us to measure the  abundance distribution of light/odd-Z (Li, Na, Al, Sc, V), $\alpha$ (Mg, Si, Ca, Ti), Fe-peak (Cr, Mn, Fe, Co, Ni, Zn), and neutron capture (Sr, Y, Zr, Ba, La, Nd, Eu) elements along the \sname\ stream. We find that the stream is (1) near solar metallicity with \feh\ = --0.03~dex and (2) has a  metallicity spread of 0.07~dex (or 0.04 dex when outliers are excluded). We also find that (3) the abundance of Li indicates that Psc--Eri is $\sim$120~Myr old, 
consistent with its gyrochronology age. Additionally, \sname\ has (4) [X/Fe] abundance spreads which are just larger than the typical uncertainty in most elements, (5) it is a cylindrical-like system whose outer edges rotate about the center and,  (6) no significant abundance gradients along its major axis except a potentially weak gradient in [Si/Fe]. These results show that \sname\ is a uniquely close, young, chemically interesting laboratory for testing our understanding of star and planet formation. 
\end{abstract}



\section{Introduction}\label{sec:intro}
Stellar clusters and associations provide crucial environments for exploring the formation and early evolution of stars and planets. 
The advent of large astrometric surveys, such as the \textit{Gaia} mission \citep{Gaiasummary2018}, 
and spectroscopic surveys, such as the Large Sky Area Multi-Object Fiber Spectroscopic Telescope \citep[LAMOST,][]{Cui2012, Luo2015} survey, among others, have enabled the discovery of new stellar clusters and associations \citep[e.g.][]{Kounkel2019}. 

Recently, \cite{Meingast2019b} discovered a 400-pc-long stellar stream in the Galactic disk with the 6D position and velocity data from the second data release from the  \gaia\ mission. Referred to as Pisces--Eridanus (\sname) by \citet{Curtis2019}\footnote{We note that \cite{Ratzenbock2020} refer to the stream as Meingast~1 in recognition of the lead author of the discovery paper.}, 
this stellar stream provides an important new source of young 
stars to study star formation and test theories of chemical and dynamical evolution of stellar systems.  This is especially true given its proximity ($d$ $\simeq$ 80-226 pc with a median distance of 129~pc)  and population size.  Initially $\sim$250 members of the stream were discovered in the stream \citep{Meingast2019b}. However, more recent studies, which used the initial 250 members as a starting point, have used a variety of techniques and uncovered upwards of 2000 members \citep[e.g.][]{Curtis2019,Ratzenbock2020, Roser2020}. For example, the recent census by \cite{Roser2020} identified 1387 members distributed across 700~pc.

\cite{Meingast2019b} noted that this newly discovered cigar-shaped stream was built up from a mass of $\sim$2000\Msun. The tightness of the main sequence shown in the color--magnitude diagram (CMD) of this system indicates that it is a single coeval stellar population. The identification of the evolved star 42~Ceti as a member indicated that it was consistent with being $\sim$1~Gyr old. Using 27 stars in common with the low resolution LAMOST spectroscopic survey, these authors found that the metallicity of the stream was \feh~=~--0.04 $\pm$~0.15~dex, consistent with a disk-like population. The results outlined in their work were interesting because it is rare to find a 1-Gyr-old star cluster close to the Sun \citep[e.g.][]{Dias2002, Kharchenko2005}. This made the \sname\ stream uniquely interesting as it would be one of the oldest structures within a 200~pc volume from the Sun. In fact, based on shape and age of \sname\ stream, \cite{Meingast2019b} concluded that that it could be a descendant of a tidally disrupting cluster or OB association. They further suggested that this structure can be thought of as a counterpart of the stellar streams that have been discovered in the stellar halo.  

Using data from NASA's \textit{Transiting Exoplanet Survey Satellite} ({\it TESS}), \citet{Curtis2019} found that the rotation period distribution of 101 \sname\ stream stars precisely overlapped with that of the Pleiades cluster, demonstrating that it is a coeval system with an age of $\approx$120 Myr, {\it much younger than the 1~Gyr age suggested by \cite{Meingast2019b}}. 
\citet{Curtis2019} also identified 34 high-mass stars as candidate members, which supports the young 120~Myr age.
Furthermore, the 3D velocity dispersion (1.3~\kms, though this value is limited by the \textit{Gaia} DR2 RV errors), is on the order of other known OB associations \citep[Tuc--Hor: 1.1~\kms;][]{Kraus2004}. However, whether this system is analogous to an evolved OB association, disintegrating cluster, or a resonant structure is still yet to be shown. 

Recently, \cite{Ratzenbock2020} uncovered as many as $\sim$2000 \sname\ stream stars and used them to infer the properties of the stream. They found that the it is  $\sim$110~Myr old and has a metallicity of Z~=~0.016 (i.e., [Fe/H]~$\sim$~--0.10~dex). They also noted that the total mass of the stream is closer to 770~\Msun. This result on the stream's age, while contrary to the initial discovery paper, is consistent with the gyrochronology age derived in \cite{Curtis2019}. Most recently, the youth of stream was also shown using the abundances of Li \citep{Arancibia-Silva2020}. However, one important, but largely missing, piece of information that can provide unique and important constraints to the nature and origins of the stream is its {\it detailed} chemical distribution. 

Therefore, in this work, we explore the detailed chemical nature of the \sname\ stream for the first time. The results enabled by these data allow us to constrain its chemical distribution, its age, and the level of chemical homogeneity along the stream. To facilitate this discussion, we describe the data used in this work in Section~\ref{sec:data}, including our own high-resolution spectroscopic follow--up at McDonald Observatory (Section~\ref{subsec:spectra}), and the low-resolution spectroscopy from the LAMOST survey (Section~\ref{subsec:LAMOST}). 
In Section~\ref{sec:methods}, we outline the methods that were used to determine the stellar parameters and chemical abundance for stars in the \sname\ stream. We present the results of the stellar spectral analysis and the detailed chemistry of the \sname\ stream in Section~\ref{sec:results}, and discuss these in context of recent literature in Section~\ref{sec:discussion}. We summarize our results and conclusions in Section~\ref{sec:conclusions}

\section{Data}\label{sec:data}

\subsection{High-Resolution Spectroscopic follow-up} \label{subsec:spectra}
In order to investigate the detailed chemical nature of the stream, 
we carried out high-resolution spectroscopic follow-up of \numstar\ \sname\ stream stars identified in Table A.1 of \cite{Meingast2019b}. Briefly, these authors identified probable candidates in the \sname\ stream by first selecting a high-quality kinematic subsample of stars from the second data release of the \gaia\ mission (\gaia\ DR2, see their Section~2). They then preformed a wavelet decomposition and searched for substructure using the DBSCAN clustering algorithm. While they found previously known structures, they also reported the discovery of a prominent previously unknown structure that \cite{Curtis2019} refers to as the \sname\ stream. We selected \numstar\ of the 258 probable \sname\ stream stars identified from these authors that were observable during mid 2019. The observational properties of the observed stars can be found in Table~\ref{tab:obsprops}. In addition to observing \numstar\ stream candidates, we also observed the twilight sky in order to obtain a solar spectrum. The solar spectrum will be used to quantify the level systematics of our atmospheric abundance measurements with respect to the literature \citep[e.g.][]{Asplund2005}. 

\begin{table*}
\caption{Observational Properties of Observed Targets}
\label{tab:obsprops}
\addtolength{\tabcolsep}{-1.5pt}
\begin{tabular}{llllllllllllll}

\hline\hline
\gaia\ Source ID & RA & DEC & $ \varpi$ & $ \sigma\varpi $ & RV$_{\rm{Gaia}}$ & $\sigma$RV$_{\rm{Gaia}}$ & RV & $\sigma$RV & X & Y & Z & SNR  & LAMOST\\
 & (deg) & (deg) & (mas) & (mas) & (\kms) & (\kms)&(\kms)&(\kms)&(pc)&(pc)&(pc)&pixel$^{-1}$&\\
 \hline
24667616384335872 & 36.9477 & 11.2409 & 7.58 & 0.05 & 16.93 & 0.97 & 18.65 & 0.16 & --85.78 & 35.60 & --93.11 & 84 & N \\
2429059676701547008 & 1.1081 & --9.1956 & 7.32 & 0.04 & 18.27 & 0.70 & 18.88 & 0.10 & 0.88 & 48.99 & --127.04 & 78 & N \\
2443101131678052480 & 0.4557 & --5.5078 & 11.21 & 0.06 & 14.51 & 0.26 & 15.33 & 0.26 & --1.34 & 37.08 & --80.90 & 111 & N \\
2449820419035086208 & 0.1418 & --1.0403 & 8.09 & 0.05 & 15.84 & 0.75 & 16.19 & 0.09 & --5.94 & 59.04 & --107.92 & 56 & N \\
2462582789799726336 & 29.9753 & --9.5046 & 8.00 & 0.05 & 18.25 & 1.13 & 19.70 & 0.09 & --49.48 & 9.91 & --113.83 & 83 & N \\
2491305817383938816 & 32.8987 & --4.6133 & 8.71 & 0.09 & 20.78 & 1.27 & 21.88 & 0.08 & --54.95 & 12.90 & --99.62 & 48 & N \\
2493286445846897664 & 33.9434 & --2.6090 & 9.53 & 0.04 & 20.85 & 0.96 & 22.64 & 0.21 & --53.40 & 13.32 & --88.91 & 89 & Y \\
2496200774431287424 & 37.7451 & --3.0514 & 9.61 & 0.04 & 20.54 & 0.97 & 23.37 & 0.09 & --57.24 & 8.15 & --86.19 & 56 & N \\
2513568007268649728 & 33.6967 & 2.2390 & 8.13 & 0.04 & 19.72 & 0.40 & 20.64 & 0.18 & --67.23 & 23.69 & --99.65 & 85 & N \\
2547792093390641024 & 6.9064 & 2.7075 & 8.21 & 0.04 & 13.18 & 0.40 & 13.94 & 0.09 & --22.11 & 57.27 & --104.71 & 59 & N \\
2577307864562003456 & 15.6237 & 6.9985 & 9.84 & 0.05 & 14.05 & 0.20 & 14.18 & 0.09 & --34.94 & 45.02 & --83.76 & 101 & Y \\
2623655475128272384 & 339.6370 & --6.2137 & 7.44 & 0.05 & 10.72 & 1.01 & 11.88 & 0.11 & 40.60 & 71.40 & --105.65 & 93 & N \\
2625503165763607808 & 340.8204 & --3.0518 & 6.86 & 0.06 & 12.82 & 0.11 & 13.01 & 0.11 & 37.80 & 83.02 & --112.83 & 69 & N \\
2649338761082702976 & 341.1805 & --2.8930 & 7.37 & 0.04 & 11.19 & 0.47 & 11.57 & 0.13 & 34.31 & 77.38 & --105.32 & 49 & N \\
2656650960084472320 & 342.6031 & 2.9805 & 6.71 & 0.05 & 10.52 & 0.84 & 10.31 & 0.18 & 27.21 & 95.37 & --110.43 & 82 & N \\
2659252645113464320 & 348.5839 & 3.7783 & 7.04 & 0.05 & 12.73 & 0.61 & 12.16 & 0.09 & 12.31 & 87.79 & --110.31 & 68 & Y \\
2664478280283853824 & 350.5519 & 6.3629 & 7.49 & 0.04 & 9.51 & 0.99 & 10.42 & 0.10 & 4.66 & 85.05 & --102.06 & 61 & Y \\
2702106351324960768 & 326.0432 & 9.0760 & 5.30 & 0.07 & 5.80 & 0.37 & 5.17 & 0.17 & 67.66 & 144.12 & --99.38 & 69 & Y \\
2708000004232308096 & 334.3629 & 5.0439 & 5.59 & 0.06 & 7.54 & 0.73 & 7.32 & 0.15 & 50.66 & 124.78 & --116.35 & 76 & Y \\
2708287457803120640 & 332.7996 & 5.2175 & 6.33 & 0.06 & 6.97 & 0.44 & 6.94 & 0.15 & 47.94 & 111.29 & --100.10 & 91 & Y \\
2709158614610269312 & 337.7289 & 6.1538 & 6.01 & 0.05 & 8.49 & 1.11 & 8.96 & 0.12 & 37.56 & 116.31 & --111.75 & 79 & N \\
2712132690484051584 & 343.9694 & 7.5212 & 6.26 & 0.07 & 10.36 & 0.48 & 9.81 & 0.17 & 19.64 & 109.80 & --113.38 & 71 & N \\
2712471584878506496 & 342.9192 & 7.0082 & 6.96 & 0.04 & 8.67 & 3.30 & 9.52 & 0.29 & 20.52 & 98.61 & --101.50 & 89 & N \\
2721370344799405056 & 330.8342 & 6.8552 & 5.53 & 0.08 & 5.81 & 0.91 & 6.77 & 0.12 & 57.11 & 131.90 & --108.35 & 71 & N \\
2727676456301311744 & 334.2828 & 11.9464 & 5.34 & 0.04 & 4.35 & 1.40 & 6.34 & 0.14 & 41.38 & 145.03 & --109.16 & 75 & N \\
2735657124012552192 & 334.0121 & 15.2087 & 4.83 & 0.05 & 4.60 & 1.13 & 4.72 & 0.11 & 40.10 & 167.29 & --113.07 & 77 & N \\
2738999879879546752 & 2.1053 & 2.5151 & 7.10 & 0.12 & 11.14 & 0.57 & 11.45 & 0.22 & --15.16 & 71.50 & --119.72 & 91 & Y \\
2739498199165473920 & 359.6010 & 2.2980 & 8.32 & 0.05 & 15.49 & 0.53 & 16.21 & 0.16 & --8.16 & 63.10 & --101.49 & 70 & Y \\
2741665405303366912 & 1.1806 & 4.2896 & 8.30 & 0.08 & 12.17 & 0.55 & 13.45 & 0.16 & --13.07 & 64.77 & --100.28 & 97 & N \\
2745159000421889792 & 0.4268 & 5.8444 & 8.27 & 0.04 & 14.38 & 0.42 & 15.68 & 0.10 & --13.29 & 67.99 & --98.57 & 82 & N \\
2760973997717658624 & 352.1613 & 8.6942 & 7.52 & 0.05 & 11.51 & 0.53 & 12.78 & 0.12 & --1.21 & 87.06 & --99.96 & 62 & N \\
2762554339524189952 & 352.1526 & 10.4095 & 7.16 & 0.05 & 10.70 & 1.86 & 10.89 & 0.11 & --3.28 & 94.06 & --102.45 & 70 & Y \\
2814781829038090752 & 345.8403 & 12.4899 & 6.05 & 0.05 & 7.26 & 1.69 & 9.32 & 0.09 & 8.55 & 120.96 & --111.08 & 66 & Y \\
3187477818011309568 & 71.9923 & --7.8237 & 5.82 & 0.03 & 20.25 & 0.85 & 22.14 & 0.11 & --132.32 & --63.17 & --87.96 & 60 & N \\
3187547465200970368 & 71.5512 & --7.5401 & 5.72 & 0.03 & 21.83 & 0.48 & 22.19 & 0.13 & --134.91 & --62.89 & --90.21 & 67 & N \\
3193528950192619648 & 59.2665 & --10.2336 & 7.04 & 0.04 & 21.55 & 0.75 & 21.99 & 0.11 & --96.32 & --36.87 & --96.95 & 64 & N \\
3197753548744455168 & 68.4807 & --8.3244 & 6.04 & 0.04 & 21.60 & 0.50 & 21.92 & 0.23 & --124.14 & --55.70 & --92.75 & 87 & N \\
3198123568764627456 & 68.8099 & --8.2592 & 6.06 & 0.03 & 20.83 & 0.73 & 21.86 & 0.09 & --124.09 & --55.97 & --91.69 & 66 & N \\
3198734278756825856 & 66.6129 & --7.6610 & 6.24 & 0.04 & 20.11 & 0.72 & 22.64 & 0.09 & --119.90 & --49.43 & --92.89 & 66 & Y \\
5164508202742806784 & 51.9693 & --10.6436 & 8.03 & 0.04 & 21.09 & 0.39 & 23.09 & 0.10 & --76.95 & --22.73 & --94.68 & 74 & N \\
5179037454333642240 & 39.7954 & --5.5396 & 8.35 & 0.04 & 19.94 & 0.77 & 20.69 & 0.10 & --65.86 & 2.67 & --99.53 & 63 & N \\
5179904664065847040 & 47.2653 & --7.0655 & 8.31 & 0.03 & 21.07 & 0.92 & 22.97 & 0.10 & --73.68 & --10.00 & --94.16 & 90 & Y \\
\hline\hline
\end{tabular}
\raggedright
NOTE: The \gaia\ DR2 source identifier of each star is given in column~1 with equatorial ICRS coordinates in columns~2 and 3. 
The parallax and its uncertainty from \gaia\ DR2 is listed in column~4 and 5, respectively. The radial velocity (RV) information from \gaia\ and its uncertainty, along with our measured RVs and its uncertainty from the high-resolution spectra obtained in this work are listed in columns 6-9, respectively. We also add the Galactic Cartesian X, Y, and Z obtained in \cite{Meingast2019b} in columns 10, 11, and 12, respectively. The signal-to-noise ratios (SNR), measured in at the continuum level at $\sim$5350~\AA, of our optical spectra are listed in column 13.  We also note in column 14 whether each observed star has also been observed in the LAMOST survey (`Y') or not (`N').
\end{table*}

The spectroscopic follow up was completed using the Tull Echelle Spectrograph \citep{Tull1995} on the 2.7m Harlan J. Smith Telescope at McDonald Observatory in mid 2019. The instrument was set up using the slit\#4 mode, which enabled us to obtain optical spectra with a resolving power of R $=\lambda/\Delta\lambda \sim$ 60000. The spectra have a wavelength coverage of $\sim$3800--9000~\AA, with increasing inter-order gaps as one moves towards the red part of the spectrum. The spectra were reduced and extracted using the standard procedures (e.g. subtraction of the bias, dividing by the flat field, optimal spectra extraction and scattered light subtraction) using the echelle package within the IRAF suite of software.\footnote{IRAF is distributed by the National Optical Astronomy Observatory, which is operated by the Association of Universities for Research in Astronomy (AURA) under a cooperative agreement with the National Science Foundation.} In order to ensure that we can measure high-quality stellar parameters and abundances, we obtained high signal-to-noise ratio (SNR) data, with SNR $>$ 50~pixel$^{-1}$ for all targets, including the solar spectrum. The approximate SNR for each star can be found in column~13 in Table~\ref{tab:obsprops}.

\begin{figure}
	 \includegraphics[width=1\columnwidth]{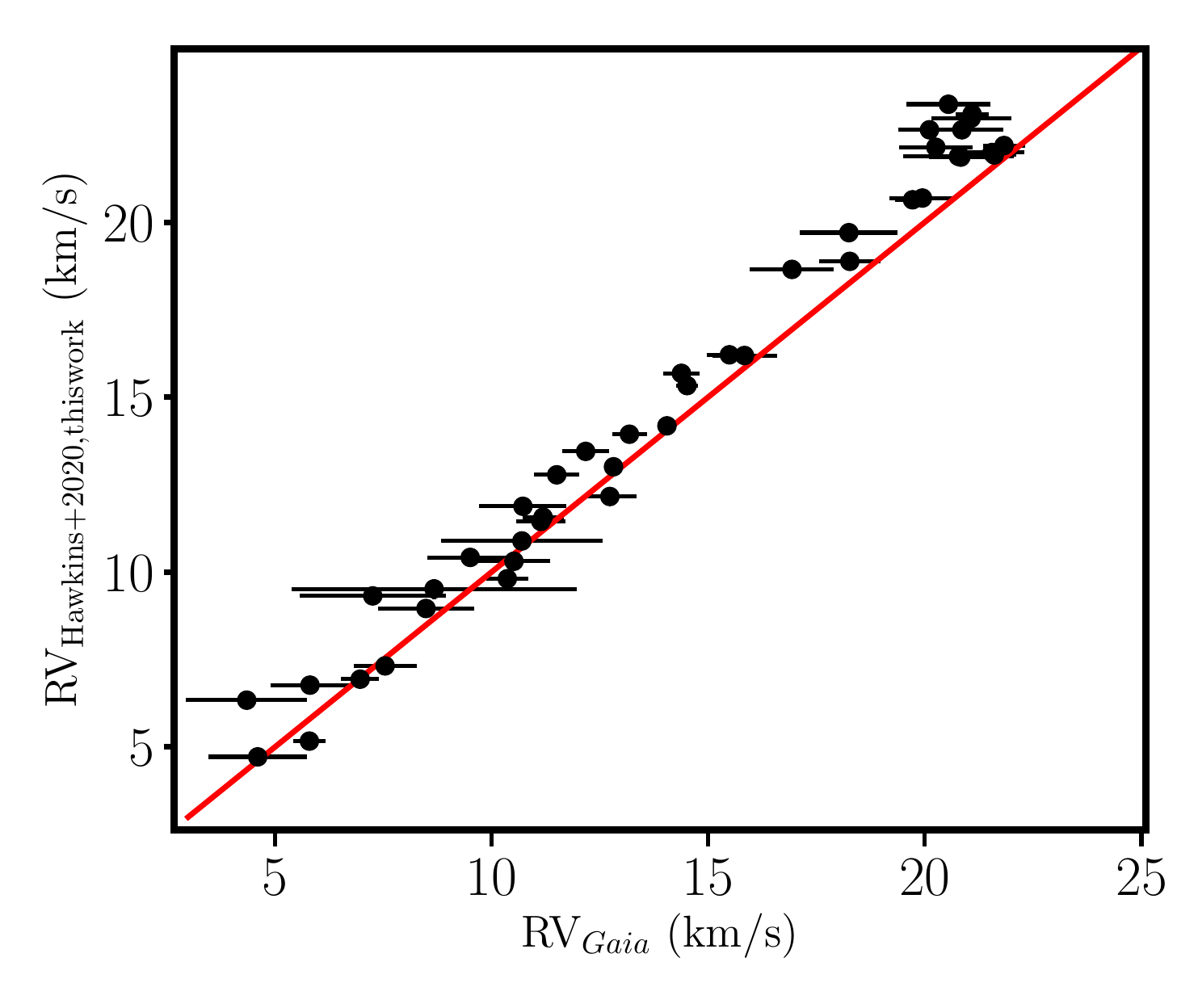}
	\caption{The radial velocities (RVs) that we measured using the high resolution spectroscopic follow-up as a function of the radial velocity measured by the \gaia\ collaboration. The red line represents the one-to-one (y=x) line.} 
	\label{fig:RVs}
\end{figure}

Once the spectra were reduced and extracted, an initial continuum normalization was done, order-by-order assuming a fifth order spline. Once each order was flattened they were stitched using the `scombine task' within IRAF. Radial velocities (RVs) were determined by cross correlation with a solar spectral template with the iSpec package \citep{2014A&A...569A.111B}. The RVs measured can be found in Table~\ref{tab:obsprops}. Fig.~\ref{fig:RVs} illustrates that the measured RVs from our high-resolution follow-up as a function of those derived from \gaia. This enabled us to confirm the \gaia\ RVs with an offset of 0.78~\kms\ and a dispersion of 0.82~\kms\ across the \numstar\ stars observed. With the reduced, extracted, wavelength- and RV-corrected data in hand, in Section~\ref{sec:methods}, we describe how we used these spectra to measure the atmospheric parameters and, more importantly, their detailed chemical nature.

\subsection{The LAMOST Survey} \label{subsec:LAMOST}
Since we are interested in the detailed chemical nature of the \sname\ stellar stream in this work, we cross-matched the stream candidates against several large multi-object spectroscopic surveys. These surveys have been developed to study the chemo-dynamical nature of the Milky Way's disc, bulge, and halo. Together these surveys have obtained low- (R$\sim$1800), moderate- (R$\sim$ 10000), and  high-resolution (R$\sim$40000) spectra for upwards of 10$^6$ stars. The surveys use those spectra to determine the stellar atmospheric parameters and chemical abundance, where possible. They additionally measure the RVs and thus, in combination with \textit{Gaia} astrometry, further constrain the kinematic nature of these stars

In an effort to determine what, if any, chemical information already exists for the \sname\ stream candidates, we cross-matched all 257 stream candidate stars identified in \cite{Meingast2019b} with the second data release of the GALactic Archaeology with HERMES \citep[GALAH DR2,][]{De_silva2015, Buder2018} survey, the 16\ith\ data release of the Apache Point Observatory Galactic Evolution Experiment \citep[APOGEE DR16,][]{Majewski2017}, and the fifth data release of the Large Sky Area Multi-Object Fibre Spectroscopic Telescope survey \citep[LAMOST DR5,][]{Luo2015, Xiang2017}. This was done using up to a 3$''$ search radius around each source with each survey. Of these surveys, there were only successful cross matches with the LAMOST DR5. 

The LAMOST survey has obtained low-resolution (R$\sim$1800) optical (3700~$<\lambda<$~9000~\AA) spectra for 8 million stars spread across the Milky Way. This is one of the largest collections of optical spectra of stars to date, and hence has the largest number of overlap with the \sname\ stream. The LAMOST catalog contains not only stellar classifications but also RV information in addition to basic stellar parameters (\teff, \logg, and \feh) for upwards of 5 million stars using the LAMOST stellar parameter pipeline \citep{Luo2015}.  

Most recently, \cite{Xiang2019} supplemented these stellar parameters with the derivation of 16 atmospheric abundances, [X/Fe], for 6 Million Stars apart of the fifth data release of the LAMOST survey. They derived these abundances via label transfer (through a neural net framework) using data from the GALAH and APOGEE surveys as a training set. To get the most out of the LAMOST data, we have cross matched this value-added stellar parameter and abundance catalog against the stream candidates. Unlike in \cite{Meingast2019b}, who found 27 stars in the stream that were also observed with the third data release of LAMOST, we found 40 stars observed with LAMOST DR5 that are also apart of the stream. In order to best combine our high-resolution abundance results with  LAMSOT, we observed 13 stars in the stream that were also observed in LAMOST (see the last column of Table~\ref{tab:obsprops} for more details). This is important specifically because it allows us to quantify the level of systematics between our abundances relative to those derived in \cite{Xiang2019} with the LAMOST spectra. Once correcting for these systematics, we can combine our results with LAMOST to explore the detailed chemical nature of this newly discovered stream in a more comprehensive way.

\section{Stellar Parameter and Abundance Analysis}\label{sec:methods}
The primary effort of this work is to measure the chemical nature of the \sname\ stellar stream in the Galactic disk. In order to do that, we must first measure the stellar atmospheric parameters, including the effective temperature (\teff), surface gravity (\logg), metallicity (\feh) and microturblent velocity (\vmicro) from the observed stellar spectra (see Section~\ref{sec:data}). To accomplish this, we used the  Brussels Automatic Code for Characterizing High accUracy Spectra \citep[BACCHUS;][]{Masseron2016} code.
BACCHUS requires the use of both a line list of atomic and molecular information as well as stellar atmosphere  models. Similar to \cite{Hawkins2019}, we take the atomic line list from the fifth version of the Gaia-ESO linelist (Heiter et al., submitted).  Hyperfine structure splitting is included for Sc I, V I Mn I, Co I, Cu I, Ba II, Eu II, La II, Pr II, Nd II, Sm II (Heiter et al., submitted).  Molecular information for the following species were also included: CH \citep{Masseron2014}, and CN, NH, OH, MgH and  C$_{2}$ (T. Masseron, private communication). SiH molecules are adopted from the Kurucz linelists\footnote{http://kurucz.harvard.edu/linelists/linesmol/} and those from TiO, ZrO, FeH, CaH from B. Plez (private communication) are also included. BACCHUS also uses the MARCS model atmosphere grid \citep{Gustafsson2008}, along with TURBOSPECTRUM \citep{Alvarez1998, Plez2012} in order to synthesize spectra.

For a detailed discussion on how BACCHUS is used to derive the stellar parameters and atmospheric abundances, we refer the reader to Section~3 of \cite{Hawkins2019}. We use the same setup, input physics (e.g., line list, model atmospheres, and line selection) as that work. Briefly, BACCHUS determines the stellar parameters using the Fe excitation-ionization balance technique. Under this technique, one tunes the \teff, \logg, and \vmicro\ of the model atmosphere in order to ensure that there is no correlations between the derived Fe abundance and the excitation potential and reduced equivalent width of the $\sim$100 lines used. In addition, the \ion{Fe}{I} and \ion{Fe}{II} abundances are required to match. 

Once the stellar parameters were determined, the individual abundances for \numel\ elements were derived. These elements span different families including the light/odd-Z (Li, Na, Al, Sc, V), $\alpha$ (Mg, Si, Ca, Ti), Fe-peak (Cr, Mn, Fe, Co, Ni, Zn), and neutron capture (Sr, Y, Zr, Ba, La, Nd, Eu). The reported elements were determined via $\chi^2$ minimization between a set of synthetic and observed spectra using the `abund' module within BACCHUS. This module starts by fixing the stellar parameters to the values derived above, then proceeds to recompute synthetic spectra for a given set of elemental lines differing the [X/Fe]\footnote{Chemical abundances are represented in the standard way as a logarithmic ratio of element X to element Y, relative to the Sun, [X/Y], such that  $\textup{[X/Y]} = \mathrm{log}\left ( \frac{N_{X}}{N_{Y}} \right )_{star} - \mathrm{log}\left ( \frac{N_{X}}{N_{Y}} \right )_{Sun}$, where $N_{X}$ and $N_{Y}$ are the number of element X and element Y per unit volume respectively. We also note that instead of assuming the solar abundances from literature \citep[e.g.][]{Asplund2005}, we derive them directly from our Twilight spectrum.}, between --0.6 $<$ [X/Fe] $<$ +0.6 and determine a best fit value by $\chi^2$ minimization. If the value of [X/Fe] is outside of this range BACCHUS will automatically adjust the range higher or lower and refit the spectra. The reported elemental abundance ratio is the median [X/Fe] value for all quality unflagged lines\footnote{BACCHUS has a decision tree algorithm which flags low-quality lines that are saturated, have too low signal-to-noise ratios, or are suspiciously high (or low). For more details the reader should consult \cite{Hawkins2015b} and \cite{Masseron2016}.}. The reported uncertainty in [X/Fe] is estimated as  the standard error in the mean for all lines used in the abundance determination. When only one absorption line is useable, we conservatively assume the uncertainty is 0.10~dex. We note that this is likely an underestimate of the total internal uncertainty as it does not propagate the uncertainty of the stellar parameters (\teff, \logg, \vmicro) to the chemical abundances. For these types of stars, it is expected this will be on the order of 0.05--0.10~dex depending on the element \citep[e.g.][]{Hawkins2019}.

 \begin{figure}
	 \includegraphics[width=1\columnwidth]{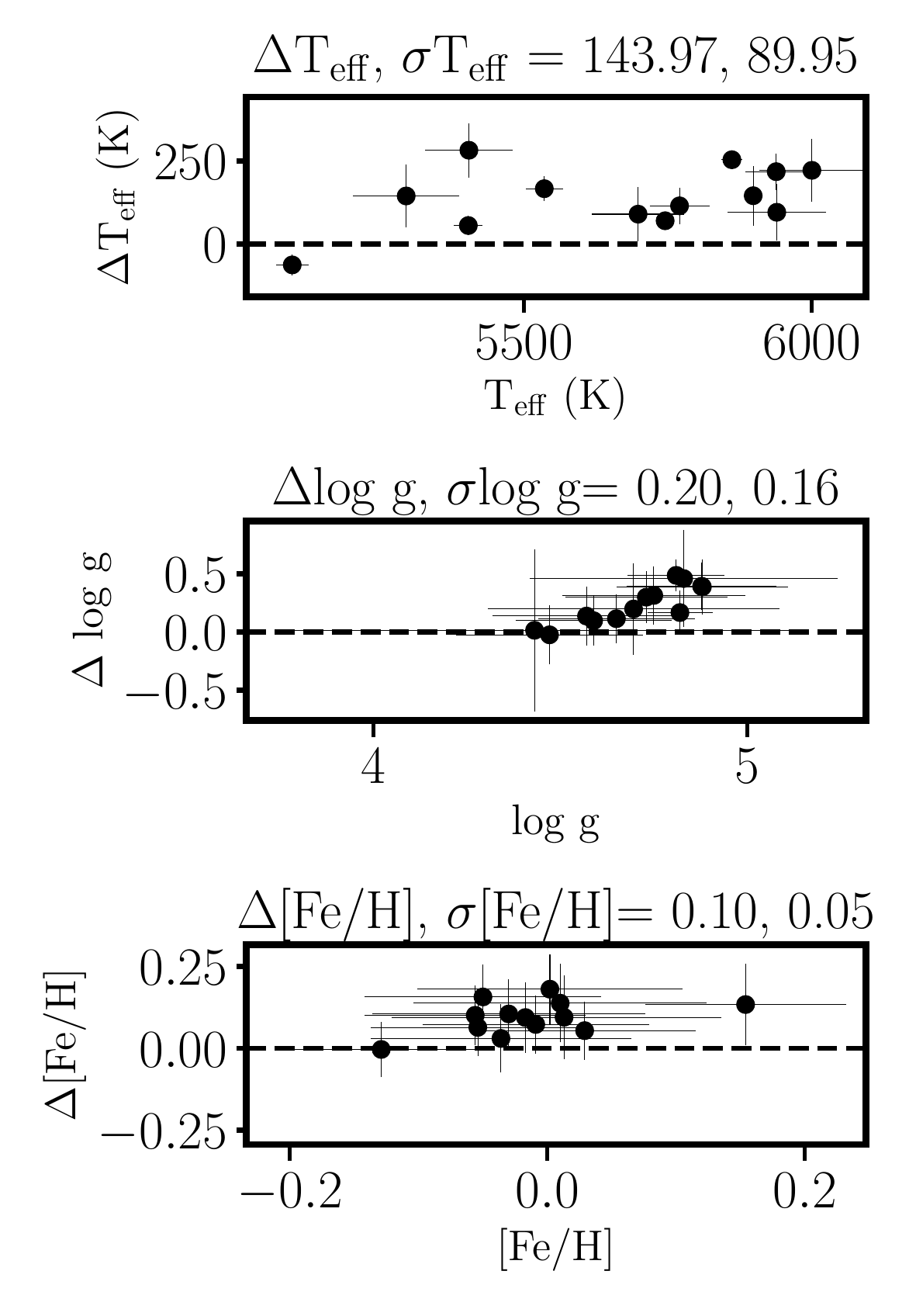}
	\caption{The difference between the LAMOST-derived \protect\citep[see][for more details]{Xiang2019} \teff\ (top panel), \logg\ (middle panel), and \feh\ (bottom panel) and the high-resolution analysis completed in this work. In this figure, the $\Delta$ denotes our values subtracted from the LAMOST values for the same stars (i.e. the LAMOST-derived \teff\ for stars in common  are $\sim$143~K smaller than our values). This includes the 13 stars where there are both LAMOST and complimentary high-resolution spectra outlined in this paper. In each panel the x-axis the high-resolution BACCHUS derived stellar parameters.} 
	\label{fig:LAMOST_diff}
\end{figure}

\section{Results}\label{sec:results}
\subsection{Stellar Parameters and Metallicity Distribution of the \sname\ Stream}
Before we determined the stellar parameters (\teff, \logg, \feh, and \vmicro) for all stars in our \sname\ stream sample, we started with deriving these parameters for our twilight spectrum of the Sun. This first test allows us to ensure that the method, selected Fe lines, and input physics (e.g., the atomic and molecular line list and model atmosphere grid) are appropriate. We report our derived Solar atmospheric parameters and detailed abundances in Table~\ref{tab:solar} using the observed twilight spectrum. In addition, we also tabulate reference values for the solar parameters and abundances taken from \cite{Asplund2005}. As shown in Table~\ref{tab:solar}, we find the following Solar atmospheric parameters:  \teff\ = 5707 $\pm$25 K, \logg\ = 4.45 $\pm$0.15, \feh\ = 0.00 $\pm$0.07,  and \vmicro\ = 0.88 $\pm$0.05~\kms. These values are encouragingly similar to the reference values of \teff\ = 5777 K, \logg\ = 4.44, \feh\ = 0.00 $\pm$0.07,  and \vmicro\ = 1.20 $\pm$0.50~\kms\ \citep[e.g.,][]{Jofre2014}. We also derived the absolute Solar atmospheric abundances\footnote{We note that the absolute abundances of species $X$, denoted as $\log{A_X}$, are in the usual form where $\log{A_X} = \log{\frac{N_X}{N_H}}$ where $\log{N_H}$ is normalized to 12.00.} for the elements studied here. The Solar abundances differ from the literature by as much 0.30~dex. In order to maintain internal consistency, we choose to use our derived solar abundances as the reference point for all [X/Fe] ratios instead of assuming them from the literature \cite[e.g.][]{Asplund2005}. 

\begin{table}
\caption{Solar Atmospheric Parameters and Chemical Abundances}
\label{tab:solar}
\addtolength{\tabcolsep}{-1pt}
\begin{tabular}{llll l l }
\hline\hline
Parameter & Value & $\sigma$Value & Lines Used & Ref Value$^a$ & $\sigma$Ref Value$^a$\\
\hline
\teff\ (K) & 5707 & 25 & ... & 5777 & 3 \\
\logg & 4.45 & 0.15 & ... & 4.437 & ... \\
\feh & 0.00 & 0.07 &109  & 0.00 & 0.05\\
\vmicro\ (\kms) & 0.88 & 0.05 & ... & ...$^b$ & ...$^b$\\
Na & 6.19 & 0.03 & 9 & 6.17 & 0.04 \\
Mg & 7.38 & 0.03 & 5 & 7.53 &0.09 \\
Al & 6.38 & 0.10 & 1 & 6.37 & 0.06 \\
Si & 7.50 & 0.04 & 13 & 7.51 & 0.04  \\
Ca & 6.32 & 0.02 & 21 & 6.31&0.04 \\
Ti & 4.80 & 0.01 & 49 & 4.90 & 0.06 \\
V & 3.82 & 0.02 & 24 & 4.00 & 0.02\\
Sc & 3.16 & 0.05 & 11 & 3.05 & 0.08 \\
Cr & 5.54 & 0.02 & 20 & 5.64&0.10 \\
Mn & 5.24 & 0.03 & 12  & 5.39 & 0.03\\
Co & 4.77 & 0.04 & 17 & 4.92 & 0.08 \\
Ni & 6.19 & 0.04 & 22 & 6.23 & 0.04\\
Cu & 4.00 & 0.06 & 5 & 4.21&0.04\\
Zn & 4.47 & 0.07 & 2 & 4.60 & 0.03\\
Cr & 5.54 & 0.02 & 20 &5.64&0.10  \\
Sr & 2.74 & 0.14 & 2 & 2.92 & 0.05\\
Y & 1.97 & 0.05 & 9 & 2.21 & 0.02\\
Zr & 2.71 & 0.05 & 6&2.59&0.04 \\
Ba & 2.23 & 0.06 & 5 &2.17 & 0.07  \\
La & 1.18 & 0.05 & 13  &1.13 & 0.05\\
Nd & 1.57 & 0.07 & 10 & 1.45 & 0.05 \\
Eu & ... & ... & 0 & 0.52 & 0.06 \\
Fe & 7.45 & 0.01 & 109 & 7.45 & 0.05 \\
\hline\hline
\end{tabular}
\raggedright

NOTE: The spectroscopically derived stellar parameters (\teff, \logg, \feh, \vmicro) and the absolute chemical abundances for the twilight spectrum of the Sun. The parameter, its derived spectroscopic value, and that parameter's uncertainty from our high resolution twilight spectrum of the Sun is tabulated in columns 1, 2, and 3, respectively. Note that the abundance tabulated here are the absolute abundance, log(X). Column~4 represents the number of lines that were used for the derivation of the solar chemical abundances. $^a$Also tabulated are the reference values and their uncertainties in column~5 and 6 taken from \cite{Asplund2005}. $^b$\vmicro\ was not used in \cite{Asplund2005}, because they use a 3-D model atmosphere of the Sun, which no longer requires the use of the \vmicro\ fudge factor \citep[see][for more details]{Asplund2000b, Asplund2005}. 
\end{table}

The atmospheric stellar parameters (\teff, \logg, \feh, and \vmicro) for all stars in our sample can be found in Table~\ref{tab:SPs}. In addition, the derived chemical abundance ratios, [X/Fe], for light/odd-Z (Li, Na, Al, Sc, V), $\alpha$ (Mg, Si, Ca, Ti), Fe-peak (Cr, Mn, Fe, Co, Ni, Zn), and neutron capture (Sr, Y, Zr, Ba, La, Nd, Eu) elements can also be found in Table~\ref{tab:SPs}. We again remind the reader that the [X/Fe] ratios use the Solar abundances derived in Table~\ref{tab:solar}. 

With these atmospheric parameters and abundances derived using high-resolution spectra in hand, we compare their values to those derived with low-resolution LAMOST spectra \citep{Xiang2019}. This was done for the 13 stars which have both high-resolution and LAMOST spectra and was done in order to put the LAMOST abundances on our scale. In Fig.~\ref{fig:LAMOST_diff}, we show the difference of the \teff\ (top panel), \logg\ (middle panel), and \feh\ (bottom panel) derived using our high-resolution spectra data with those values determined using the low-resolution LAMOST spectra. We note that the $\Delta$ is defined as our high-resolution derived values minus those from LAMOST. Also displayed are the mean and dispersion of the difference between our high-resolution analysis and the low-resolution derived values from LAMOST. For \teff\ we find an offset of  144~K with a dispersion of 90~K in \teff, 0.20~dex with a dispersion of 0.18~dex in \logg, and 0.10~dex with a dispersion of 0.05~dex in \feh. The mean offset gives us an estimate of the systematic differences between LAMOST and our high resolution analysis while the dispersions provide an estimate of the precision. These offsets and dispersion are not uncommon when comparing high-resolution to low-resolution data sets \citep[e.g.][]{2008AJ....136.2022L,Boeche2018} and are only used to zero-point shift all LAMOST [X/Fe] abundance ratios onto our stellar abundance scale. For metallicity,  this zero-point shift amounts to adding +0.10~dex to the \feh\ derived from LAMOST spectra by \cite{Xiang2019}. 

In Fig.~\ref{fig:metdist}, we show the metallicity distribution for the \numstar\  \sname\ stars observed in this work (black line). For completeness, we also show the metallicity distribution for the 40 \sname\ stars observed by LAMOST with an additive 0.10~dex systematic offset in order to ensure the LAMOST \feh\ are on our metallicity scale. With the \numstar\  \sname\ stars observed in this work, we find a metallicity distribution that has a peak at --0.03~dex with a dispersion of 0.07~dex. While this confirms the stream metallicity found in \cite{Meingast2019b}, which showed that the peak \feh\ was  --0.04~dex with 27 stars, we find a significantly lower dispersion by a factor of $\sim$2. While the peak of the metallicity distribution of the stream is found to be --0.03~dex using the LAMOST values (as expected), the dispersion is significantly larger ($\sim$0.15~dex), this is expected due to the lower resolution spectra and thus larger uncertainties with which \feh\ can be measured. 

\begin{figure*}
	 \includegraphics[width=2\columnwidth]{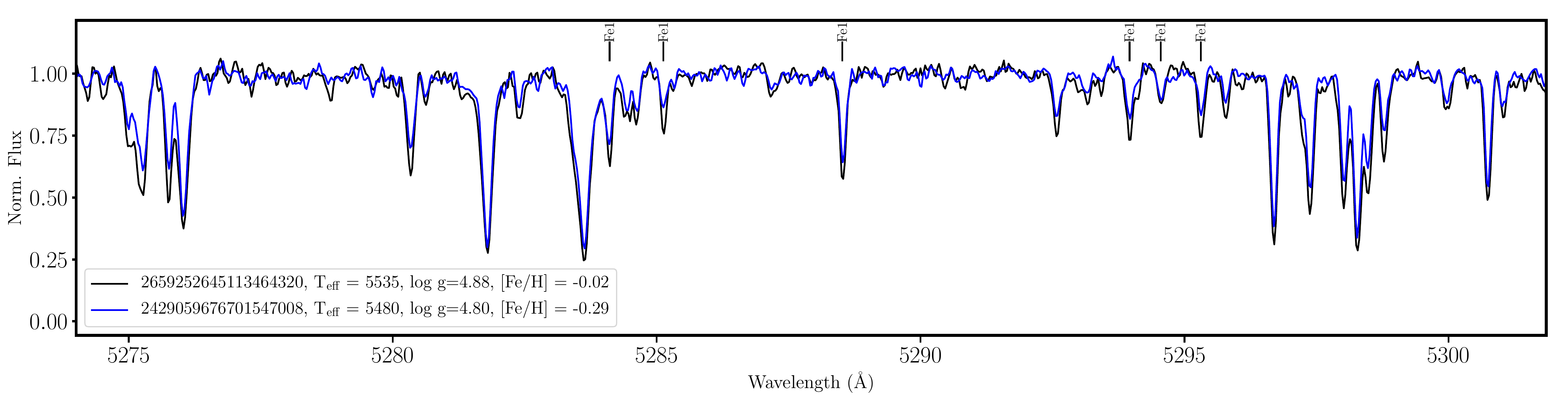}
	\caption{Illustrates the spectral differences between \gaia\ DR2~2429059676701547008 (the most metal-poor star we uncovered in \sname) and a comparison star \gaia\ DR2~2659252645113464320 in the 5274--5305~\AA\ wavelength regime. The comparison star was chosen to have very similar \teff\ and \logg\ as \gaia\ DR2~2429059676701547008 but with a metallicity that is close to the median \feh\ of \sname. }
	\label{fig:spec}
\end{figure*}

\begin{figure}
	 \includegraphics[width=1\columnwidth]{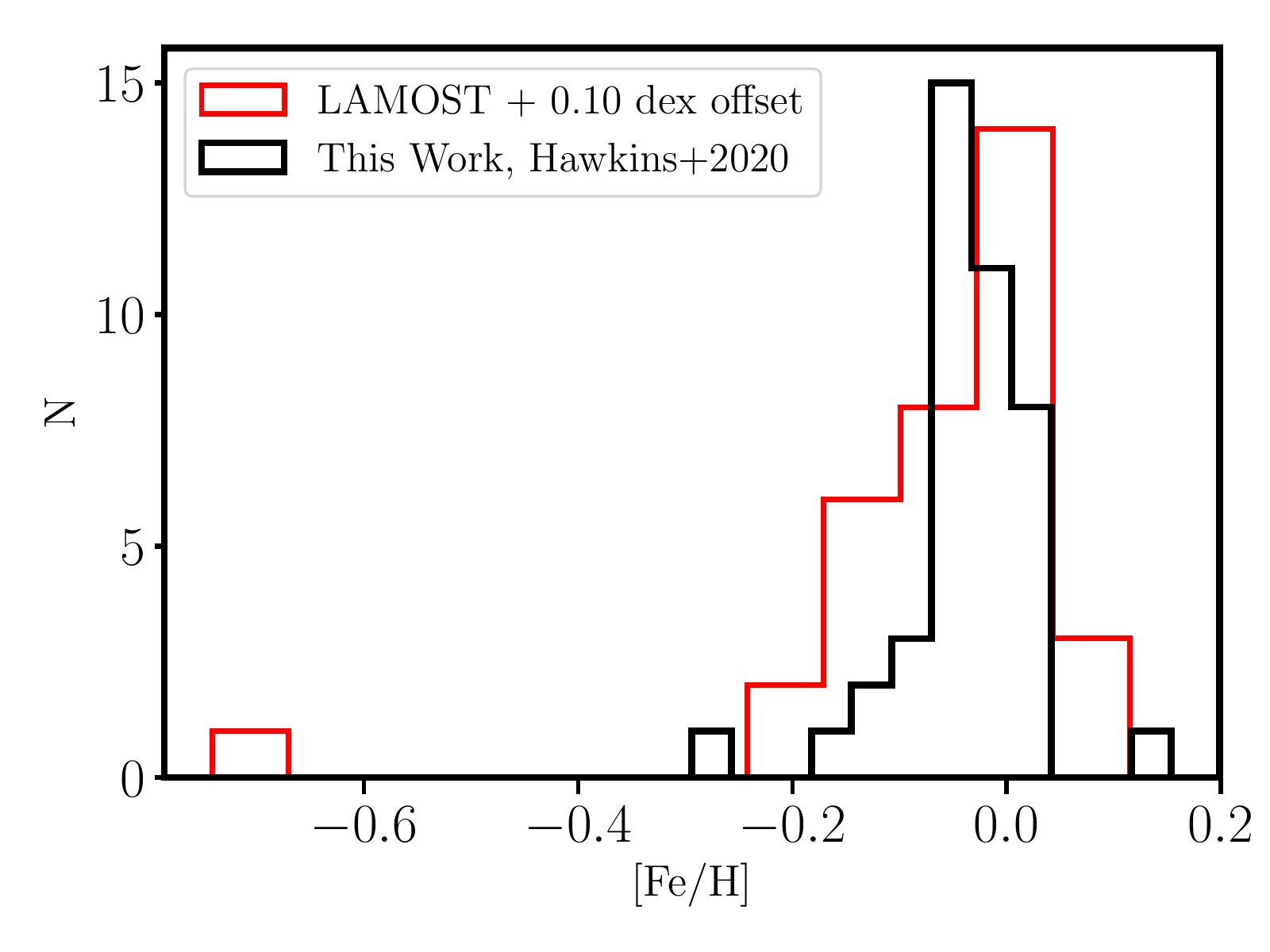}
	\caption{The metallicity distribution of the \sname\ stream as measured by our high-resolution follow-up of \numstar\ stars (black). Also shown is the metallicity distribution of the \sname\ stream as measured by \protect\cite{Xiang2019} on the 40 stars in common with the LAMOST survey (red). An offset of 0.10~dex is applied to the LAMOST metallicity and is determined by the bias that results from comparing the \feh\ between LAMOST and our high-resolution results. The measured median \feh\ for the \sname\ stream is --0.03~dex with a dispersion of 0.07~dex. } 
	\label{fig:metdist}
\end{figure}

The dispersion in \feh\ appears to be driven by a number of outlier stars, namely \gaia\ DR2 2429059676701547008, which has a \feh\ = $-$0.29~dex (3$\sigma$ outlier), and \gaia\ DR2~2577307864562003456, which has a metallicity of +0.15~dex (2.5$\sigma$ outlier). These two stars have high-quality astrometric and kinematic information (with parallax uncertainties well below 1\%, and RVs that are consistent between \gaia\ and our high resolution analysis). In Fig.~\ref{fig:spec} we illustrate the spectral difference between the metal-poor outlier star \gaia\ DR2~2429059676701547008 and a comparison star with very similar \teff\ and \logg\ but whose \feh\ is more metal-rich near the median value for \sname. This figure clearly shows that \gaia\ DR2~2429059676701547008 is significantly more metal-poor (with significantly weaker absorption features) compared to the typical metallicity found in \sname. 
We have no reason to believe that either it or \gaia\ DR2~2577307864562003456 are non-members based 
on their kinematics and spatial positions; therefore, we choose to keep both stars in this analysis.\footnote{However, we note that the metal-poor star (\gaia\ DR2~2429059676701547008) appears to be chromospherically less active (according to the \caiihk, H$\alpha$, and Ca IR triplet lines) than the reference star (\gaia\ DR2~26592526451134643206), 
and it even appears to be as inactive as old field stars with similar stellar properties. 
Until we have systematically investigated magnetic activity in Psc--Eri, and compared it to the well-studied Pleiades to establish the typical behavior for bona fide members, 
it is premature to rule out membership based on inactivity alone. Unfortunately, the metal-poor star was not observed by \textit{TESS}, meaning we cannot measure its rotation period, which could also corroborate its youthful membership of Psc--Eri.}

We note that \gaia\ DR2~2577307864562003456 is considered a member by both \cite{Meingast2019b} and  \cite{Roser2020}, where the latter authors claim to do a more conservative kinematic selection of the stream. However, \gaia\ DR2~2429059676701547008 is only considered a member by \cite{Meingast2019b} but is not in \cite{Roser2020}\footnote{While \cite{Roser2020} does not claim that \gaia\ DR2~2429059676701547008 is a member of the \sname\ stream they also do not rule it out as a member.}. Removing \gaia\ DR2~2429059676701547008 would lead to a reduction in the overall \feh\ dispersion to 0.05~dex. If we only select those \sname\ stars that we observed, which have been identified as members in {\it both} \cite{Roser2020} and \cite{Meingast2019b}, we would find an overall \feh\ dispersion of 0.04~dex with 29 total stars. Additionally, we note that we find a very small systematic trend between \teff\ and \feh, which when corrected for, yields a metallicity distribution that is symmetric with a dispersion of 0.035~dex for these 29 stars. Together these indicate that it is likely that the actual stream has a $\sigma$\feh $\leq$ 0.04~dex, but larger studies are required to further identify the nature of the outlier stars. 

A similar dispersion of 0.05~dex was measured for 20 members of the Pleiades by \citet{Soderblom2009}, using Keck/HIRES spectra analyzed with Spectroscopy Made Easy \citep[SME;][]{sme, Valenti2005}.
Magnetic activity is known to affect absorption line profiles \citep[e.g., see figure 1 in][]{Davis2017}, which hinders precision Doppler searches for exoplanets \citep{Dumusque2018}.
\citet{JYG2019} analyzed spectra for the $\sim$400-Myr-old solar twin HIP~36515 using 
an equivalent width based Fe excitation/ionization balance approach with MOOG \citep{Sneden1973}, 
and found that the derived stellar properties (\teff, \logg, \vmicro, and \feh) correlated 
with the $S_{\rm HK}$ chromospheric emission index, 
with \feh\ varying by 0.016~dex between $S_{\rm HK} = 0.301$-0.355. Our Psc--Eri targets are younger than HIP~36515, and so are likely even more active \citep[e.g.,][]{Skumanich1972,mamajek2008}; therefore, the impact of magnetic activity on our chemical abundances might be comparable or even larger.

The above discussion illustrates that (1) a careful revisit of \sname\ membership and (2) a larger study of the metallicity distribution of the \sname\ stream will be required in the future to confirm (or not) the existence of the \feh\ spread (i.e. to determine to what extent the chemical outliers are in fact really apart of the cluster), and (3) the impact of magnetic activity on chemical abundance measurements will need to be further investigated. The second of these will be easily facilitated with the more than 1000 \sname\ members identified by \cite{Roser2020}.

\begin{figure*}
	 \includegraphics[width=2\columnwidth]{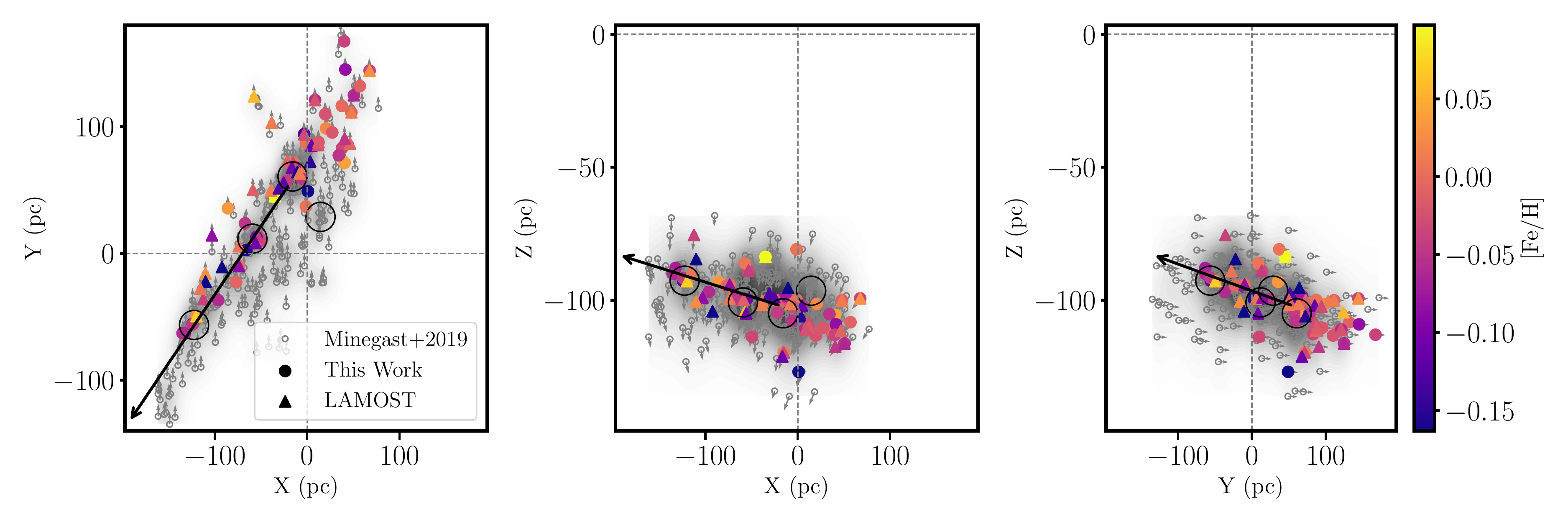}
	\caption{The Galactic X-Y (left panel), X-Z (middle panel), and Y-Z (right panel) coordinate projections of the \sname\ stream. All stream members from \protect\cite{Meingast2019b} are shown as the open gray circles. The gray density in the background is a Gaussian Kernel density estimator for these data. Stars are color-coded by metallicity where known either by LAMOST (triangles) or high-resolution (circles) spectra. The black arrows in each sub panel represent the major axis of the stream while the smaller grey arrows represent the velocity of the stars.   } 
	\label{fig:met_across}
\end{figure*}

In Fig.~\ref{fig:met_across}, we expand upon Fig.~\ref{fig:metdist}, and show how \feh\ is distributed along the stream in the X-Y (left panel), X-Z (middle panel) and Y-Z (left panel) projections. We also show in the background all \sname\ stream candidates from \cite{Meingast2019b}, and mark  possible overdensities in the stream.  For completeness, we denote the 3 dimensional (3-D) velocity (Vx, Vy, Vz) as small gray arrows in each sub panel in Fig.~\ref{fig:metdist}. The `principal axis' of the stream is also shown in each panel of Fig.~\ref{fig:metdist} as a solid black arrow. The principal axis of the stream is determined by preforming a principal component analysis (PCA) on the 3-D spatial location (in X, Y, and Z) of all stars in the stream. The first principal component is what we refer to as the `principal axis' of the stream and accounts for the largest spatial scatter. Once this is done, we perform a linear regression between each element abundance ratio and the location of the star along the principal axis of the stream. Fig.~\ref{fig:met_across} implies that there are no significant spatial metallicity gradients along the stream.

\begin{table*}
\caption{Stellar Atmospheric Parameters and Chemical Abundances of Observed Targets}
\label{tab:SPs}
\addtolength{\tabcolsep}{-1pt}
\begin{tabular}{cccccccccccccc}
\hline \hline
\gaia\ Source ID & \teff  & $\sigma$\teff & \logg & $\sigma$\logg & \feh & $\sigma$\feh & N$_{\rm{[Fe I/H]}}$ & \vmicro & $\sigma$\vmicro & [Na/Fe] & $\sigma$[Na/Fe] & N$_{\rm{[Na/Fe]}}$ & ... \\
 & (K) & (K) & (dex) & (dex) & (dex) & (dex) & (dex) & (\kms) & (\kms) & (dex) & (dex) & \\ 
\hline
2429059676701547008 & 5480 & 30 & 4.80 & 0.22 & -0.29 & 0.01 & 94 & 0.63 & 0.06 & 0.07 & 0.03 & 9 & ... \\
2462582789799726336 & 5582 & 30 & 4.70 & 0.25 & -0.03 & 0.01 & 95 & 1.14 & 0.04 & -0.02 & 0.03 & 9 & ... \\
2493286445846897664 & 5938 & 54 & 4.57 & 0.25 & -0.03 & 0.01 & 69 & 1.28 & 0.07 & 0.04 & 0.04 & 9 & ... \\
2496200774431287424 & 5641 & 30 & 4.91 & 0.28 & 0.01 & 0.01 & 87 & 1.18 & 0.05 & -0.05 & 0.03 & 7 & ... \\
2623655475128272384 & 5847 & 30 & 4.60 & 0.17 & 0.04 & 0.01 & 96 & 1.24 & 0.05 & -0.03 & 0.04 & 9 & ... \\
2625503165763607808 & 5685 & 30 & 4.57 & 0.20 & -0.03 & 0.01 & 97 & 0.98 & 0.05 & -0.02 & 0.04 & 10 & ... \\
2649338761082702976 & 5405 & 158 & 4.44 & 0.67 & -0.05 & 0.02 & 73 & 1.70 & 0.09 & -0.12 & 0.04 & 8 & ... \\
2702106351324960768 & 5698 & 80 & 4.47 & 0.25 & -0.04 & 0.01 & 84 & 1.61 & 0.06 & 0.13 & 0.07 & 9 & ... \\
2708000004232308096 & 5770 & 52 & 4.75 & 0.24 & -0.06 & 0.01 & 90 & 1.21 & 0.05 & 0.00 & 0.05 & 9 & ... \\
2708287457803120640 & 5939 & 85 & 4.65 & 0.21 & -0.01 & 0.01 & 89 & 1.28 & 0.06 & -0.01 & 0.05 & 10 & ... \\
2709158614610269312 & 5359 & 64 & 4.69 & 0.40 & -0.00 & 0.01 & 92 & 1.29 & 0.05 & -0.08 & 0.04 & 10 & ... \\
2721370344799405056 & 5679 & 42 & 4.67 & 0.28 & -0.01 & 0.01 & 91 & 1.17 & 0.05 & 0.07 & 0.04 & 12 & ... \\
2727676456301311744 & 5658 & 67 & 4.59 & 0.29 & -0.08 & 0.01 & 87 & 1.22 & 0.05 & -0.01 & 0.04 & 8 & ... \\
2735657124012552192 & 5847 & 51 & 4.54 & 0.25 & -0.04 & 0.01 & 91 & 1.23 & 0.05 & -0.04 & 0.04 & 10 & ... \\
2738999879879546752 & 5745 & 30 & 4.73 & 0.22 & 0.01 & 0.01 & 74 & 1.20 & 0.07 & -0.03 & 0.04 & 9 & ... \\
2739498199165473920 & 6000 & 92 & 4.88 & 0.23 & 0.01 & 0.01 & 80 & 1.34 & 0.08 & -0.05 & 0.04 & 10 & ... \\
2741665405303366912 & 5706 & 40 & 4.69 & 0.37 & -0.03 & 0.01 & 87 & 1.18 & 0.05 & -0.02 & 0.05 & 10 & ... \\
2745159000421889792 & 5391 & 46 & 4.82 & 0.07 & -0.07 & 0.01 & 100 & 1.17 & 0.04 & -0.03 & 0.04 & 12 & ... \\
2762554339524189952 & 5097 & 30 & 4.59 & 0.21 & -0.13 & 0.01 & 84 & 1.31 & 0.05 & -0.04 & 0.03 & 7 & ... \\
3187477818011309568 & 5143 & 30 & 4.65 & 0.47 & -0.07 & 0.01 & 88 & 1.10 & 0.05 & -0.10 & 0.04 & 8 & ... \\
3187547465200970368 & 5492 & 79 & 4.50 & 0.65 & -0.06 & 0.01 & 88 & 1.64 & 0.06 & 0.13 & 0.04 & 9 & ... \\
3197753548744455168 & 5819 & 78 & 4.44 & 0.50 & -0.12 & 0.02 & 71 & 1.86 & 0.07 & 0.12 & 0.05 & 9 & ... \\
3198734278756825856 & 5403 & 30 & 4.70 & 0.39 & 0.03 & 0.01 & 98 & 0.89 & 0.05 & -0.12 & 0.02 & 7 & ... \\
5179037454333642240 & 5083 & 86 & 4.55 & 0.87 & -0.15 & 0.02 & 88 & 1.90 & 0.06 & 0.23 & 0.04 & 11 & ... \\
5179904664065847040 & 5861 & 30 & 4.81 & 0.13 & 0.00 & 0.01 & 95 & 1.25 & 0.06 & -0.04 & 0.04 & 9 & ... \\
2664478280283853824 & 5404 & 76 & 4.83 & 0.41 & -0.05 & 0.01 & 93 & 1.62 & 0.05 & 0.07 & 0.05 & 10 & ... \\
2712132690484051584 & 5694 & 77 & 4.64 & 0.46 & -0.02 & 0.01 & 84 & 1.37 & 0.06 & 0.07 & 0.08 & 7 & ... \\
2712471584878506496 & 5864 & 60 & 4.61 & 0.41 & 0.02 & 0.02 & 61 & 0.65 & 0.08 & -0.09 & 0.07 & 5 & ... \\
2760973997717658624 & 5524 & 30 & 4.73 & 0.40 & 0.00 & 0.01 & 83 & 1.13 & 0.05 & -0.02 & 0.05 & 7 & ... \\
2814781829038090752 & 5295 & 92 & 4.43 & 0.70 & -0.05 & 0.01 & 95 & 1.61 & 0.04 & 0.12 & 0.03 & 11 & ... \\
2577307864562003456 & 5898 & 30 & 4.82 & 0.09 & 0.15 & 0.01 & 92 & 0.97 & 0.04 & -0.07 & 0.04 & 9 & ... \\
2547792093390641024 & 5009 & 40 & 4.80 & 0.41 & -0.04 & 0.01 & 86 & 1.05 & 0.06 & -0.16 & 0.04 & 7 & ... \\
2513568007268649728 & 5676 & 78 & 4.57 & 0.30 & -0.05 & 0.01 & 84 & 1.17 & 0.05 & 0.02 & 0.04 & 8 & ... \\
3193528950192619648 & 5315 & 30 & 4.80 & 0.44 & -0.06 & 0.01 & 92 & 1.11 & 0.06 & -0.11 & 0.06 & 9 & ... \\
3198123568764627456 & 5279 & 89 & 4.27 & 0.77 & -0.04 & 0.02 & 91 & 1.53 & 0.06 & 0.14 & 0.04 & 8 & ... \\
2491305817383938816 & 4906 & 30 & 4.82 & 0.49 & -0.11 & 0.02 & 82 & 1.11 & 0.06 & -0.07 & 0.05 & 7 & ... \\
24667616384335872 & 5713 & 56 & 4.75 & 0.17 & 0.04 & 0.01 & 87 & 1.19 & 0.07 & -0.09 & 0.05 & 9 & ... \\
5164508202742806784 & 5558 & 30 & 4.78 & 0.22 & -0.01 & 0.01 & 92 & 1.14 & 0.06 & -0.05 & 0.03 & 9 & ... \\
2449820419035086208 & 5048 & 102 & 4.58 & 0.84 & -0.05 & 0.01 & 93 & 1.32 & 0.05 & 0.10 & 0.03 & 11 & ... \\
2443101131678052480 & 5756 & 89 & 4.55 & 0.42 & 0.01 & 0.01 & 80 & 1.46 & 0.05 & 0.03 & 0.05 & 7 & ... \\
2659252645113464320 & 5535 & 32 & 4.88 & 0.20 & -0.02 & 0.01 & 91 & 1.08 & 0.06 & -0.01 & 0.04 & 10 & ... \\
2656650960084472320 & 5936 & 30 & 4.64 & 0.12 & -0.02 & 0.01 & 78 & 1.28 & 0.06 & -0.03 & 0.05 & 9 & ... \\

\hline \hline
\end{tabular}
\raggedright

NOTE: A cutout of a larger online table, which includes the spectroscopically derived stellar parameters (\teff, \logg, \feh, \vmicro) and the chemical abundances [X/Fe] for the \numstar\ \sname\ stream stars observed in this work. The full table will be provided in the online material. The \gaia\ DR2 source identifier of each star is given in column 1. The stellar parameters and their uncertainties (\teff, $\sigma$\teff, \logg, $\sigma$\logg, \feh, $\sigma$\feh, \vmicro, $\sigma$\vmicro) are found in columns~2-9, respectively. The chemical abundance ratio for [Na/Fe] is found in column~10 and its uncertainty in column~11. The abundance ratio is determined as the median abundance over all high quality lines of a given element (for Na the number of lines used is listed in column~12).  We note that this uncertainty is determined as the dispersion in the [Na/Fe] over all lines used to derive [Na/Fe] divided by the square root of the number of lines used.  
\end{table*}

\subsection{Lithium Depletion and the Young Age of the \sname\ Stream }
 \begin{figure}
	 \includegraphics[width=1\columnwidth]{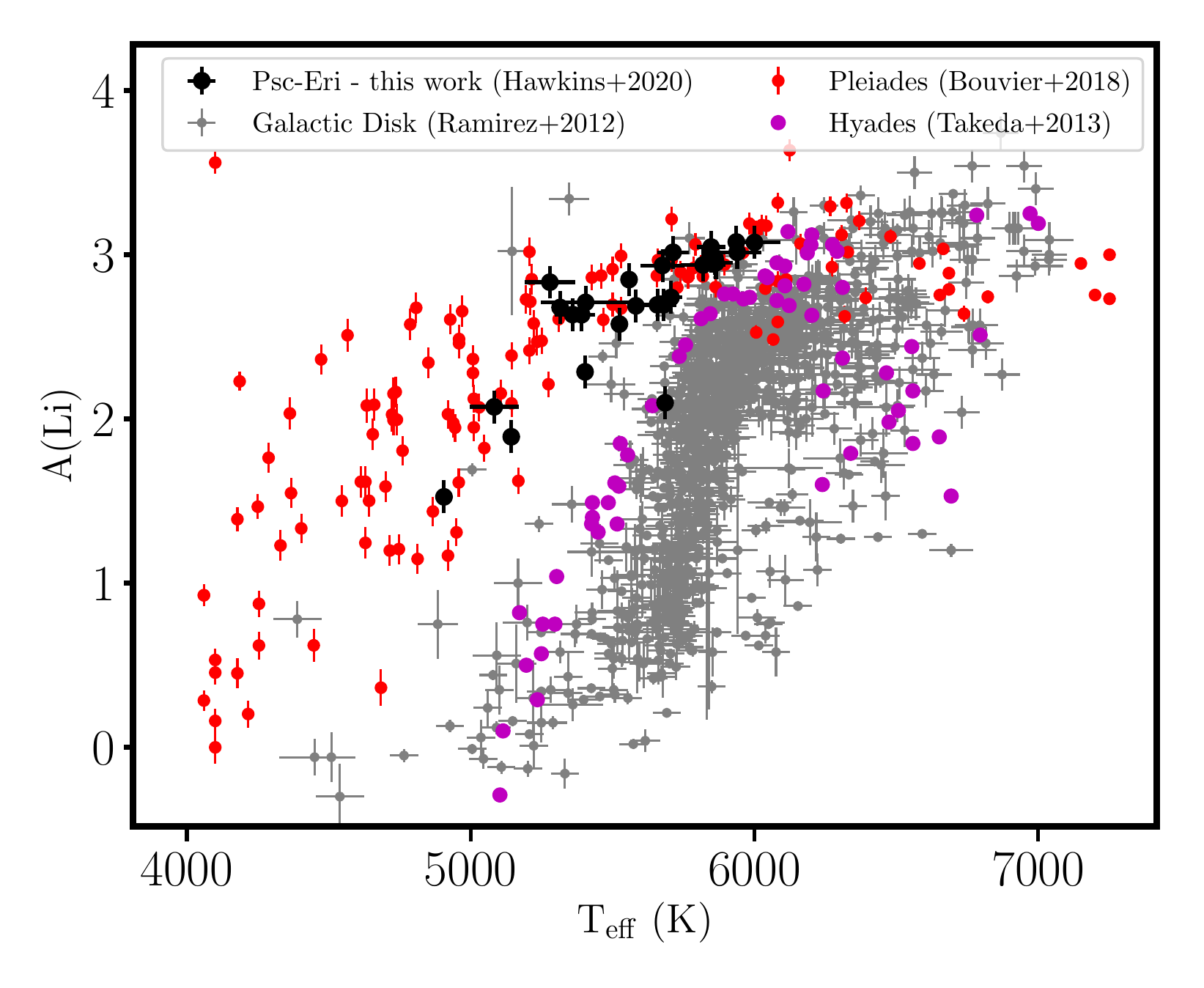}
	\caption{The measured Li abundance, A(Li) as a function of \teff, for a sample of \sname\ stream stars observed in this work (black circles). For reference, we also show the A(Li) as a function of \teff\ for stars in the Galactic Disk \citep{Ramirez2012}, the Pleiades \citep{Bouvier2018}, and the Hyades \citep{Takeda2013}. Together these indicate the \sname\ stream is of similar age as the Pleiades in the 100-200~Myr range ruling out an older 1~Gyr stream as initially suggested.    } 
	\label{fig:Li}
\end{figure}
The atmospheric abundance of lithium (Li), or lack thereof, is one of the key elements to distinguish if the stream is 1~Gyr \citep[suggested by][using fitting the color magnitude diagram]{Meingast2019b}\footnote{We note that the color magnitude isochrone analysis result laid out in \cite{Meingast2019b} was driven by the presence of the giant system "42 Ceti," which is 1 Gyr old.} or younger \citep[suggested by][using gyrochronology]{Curtis2019}.  The reason the detection of significant Li can be used as an indicator of youth is because its abundance in stellar photospheres tends to decrease over time. This happens because Li is a rather fragile element which, when mixed down into a layer in a star that is roughly 2.5 $\times 10^6$~K, is burned into heavier elements, and thus destroyed. The Li can be mixed downward either through rotationally-induced mixing, atomic diffusion, and other physical processes  \citep[e.g.][and references therein]{Michaud1986, Charbonnel2005, Pinsonneault2010, Soderblom2010, Ramirez2012, Jeffries2014}. Additionally, the strength of this depletion is expected to vary with mass (and thus \teff\ as a proxy for mass on the main-sequence). It has been shown that the abundance of Li, A(Li), as a function of \teff\ is expected to decrease with decreasing \teff\  in the Galactic disk \citep[e.g.][]{Ramirez2012}, 
among solar twins \citep[e.g.][]{Carlos2019}, 
as well in open clusters \citep[e.g.][]{Boesgaard1998, Jeffries2000, Sestito2003, Cargile2010, Cummings2012, Takeda2013, Martin2018, Bouvier2018}. 

Similar to other studies of Li at optical wavelengths, we measure its atmospheric abundance using the feature at 6708~\AA. Li was measured by synthesizing the Li feature, with different absolute Li abundances (ranging from 0--3~dex), and preforming a $\chi^2$ minimization between the synthesis and observation  using the BACCHUS `abund' module. In Fig.~\ref{fig:Li}, we show the measured absolute abundance of Li, A(Li), for the stars in the \sname\ stream as a function of the measured \teff. For reference, we also show the A(Li) as a function of \teff\ for stars in the Galactic Disk \citep{Ramirez2012}, the Pleiades \citep{Bouvier2018}, and the Hyades \citep{Takeda2013}. Fig.~\ref{fig:Li} illustrates that the A(Li) trend with \teff\ follows very closely that observed in the 120~Myr Pleiades. At a given \teff, the A(Li) within the stream is significantly higher than expected for an older cluster \citep[e.g., the Hyades;][]{Takeda2013} and thus we can rule out an older age ($\gtrapprox$200~Myr) for the stream. The A(Li) trend with respect to  \teff\  (Fig.~\ref{fig:Li}) shows that the Psc-Eri stream is about as young  as the Pleiades ($\approx$120~Myr). This result is consistent with the result from \cite{Arancibia-Silva2020} showing the Li depletion pattern most likely implies an age of 125~Myr.

\subsection{$\alpha$ elements (Mg, Si, Ca, Ti)}
 \begin{figure}
	 \includegraphics[width=1\columnwidth]{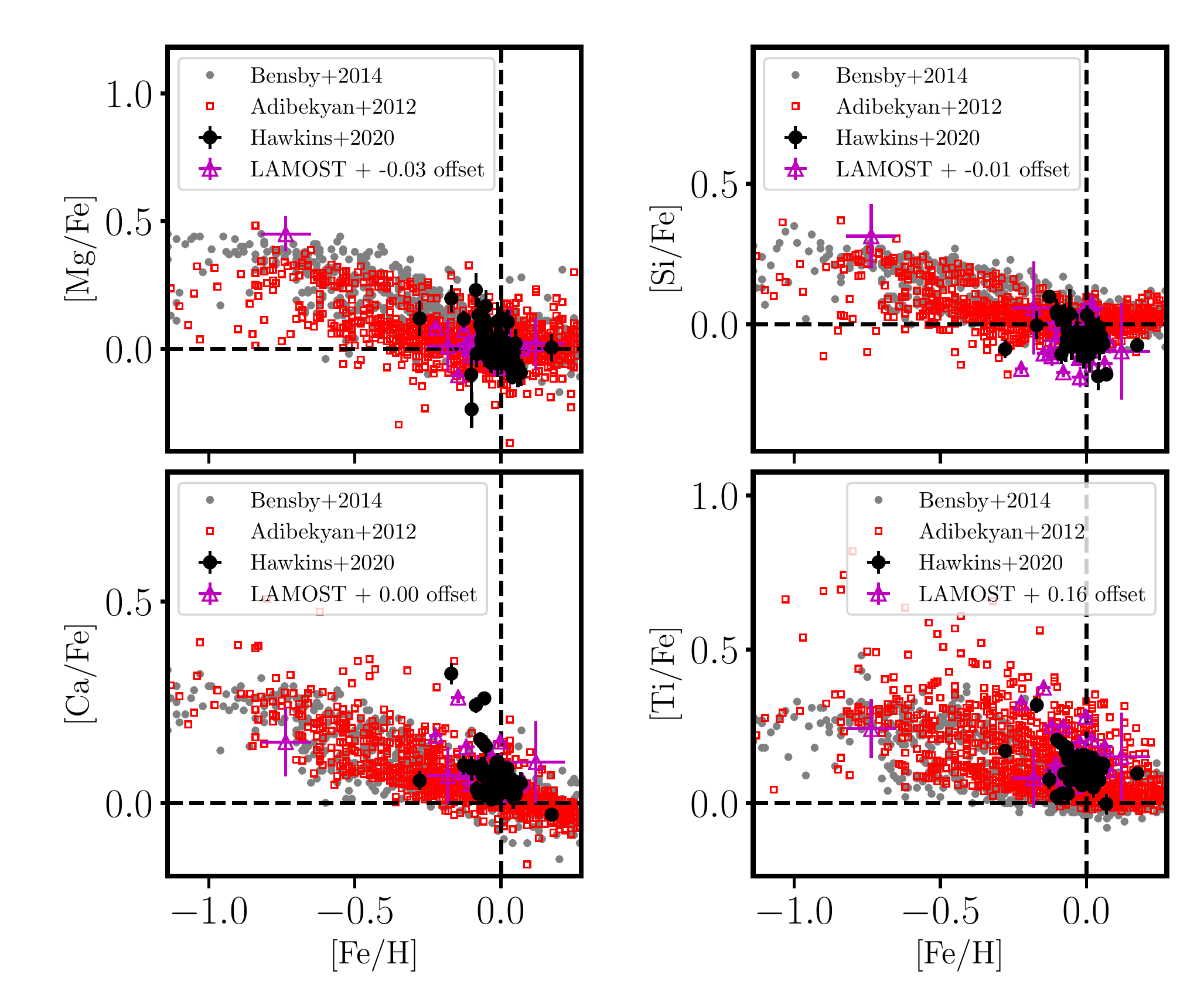}
	\caption{The atmospheric abundances ratios, [X/Fe], for the observed \sname\ stream candidates (black circles) for $\alpha$ (Mg, Si, Ca, Ti) elements as a function of metallicity. For reference, behavior of each element with respect to metallicity is also shown for the Galactic disk. The reference data for the Galactic disk are taken from \protect\citet[gray circles,][]{Bensby2014} and \protect\citet[red open squares,][]{Adibekyan2012}. } 
	\label{fig:alpha}
\end{figure}

Generally it is thought that $\alpha$ elements represent those which are formed through the successive fusion of helium nuclei (i.e. $\alpha$ particles) during the later stages of nuclear burning in the inner regions of evolved massive stars. More specifically, Mg is produced through hydrostatic burning in massive stars, while Si and Ca are predominantly produced via explosive burning during Type~II supernovae. In this work, we focus on the $\alpha$ elements: Mg, Si, Ca, and Ti. We note that Ti is often considered an $\alpha$ element because its behavior with respect to metallicity tends to be similar to the other $\alpha$ elements. However, it is sometimes classified as an Fe-peak element from a nucleosynthesis prospective. This is because it likely originates from the decay of $^{48}$Cr rather than successive capture of $\alpha$ particles, \citep[e.g.][]{Curtis2019sn}. These elements are thought to be dispersed into the interstellar medium by Type~II supernovae \citep[e.g.][and references therein]{Nomoto2013}. 

For a young population of stars ($<$5~Gyr) with near solar metallicity, as the \sname\ has been shown to be (e.g., see Fig.~\ref{fig:metdist} and Fig.~\ref{fig:Li}),  it is expected that the [Mg, Si, Ca, Ti/Fe] abundance ratios are relatively low or near zero. It might also be expected that if the stars in the \sname\ stream are co-eval and formed together, that they should all have the same chemical abundances \citep[e.g.][]{Bovy2016, Liu2016, Ness2018, Hawkins2019}. 

 \begin{figure}
	 \includegraphics[width=1\columnwidth]{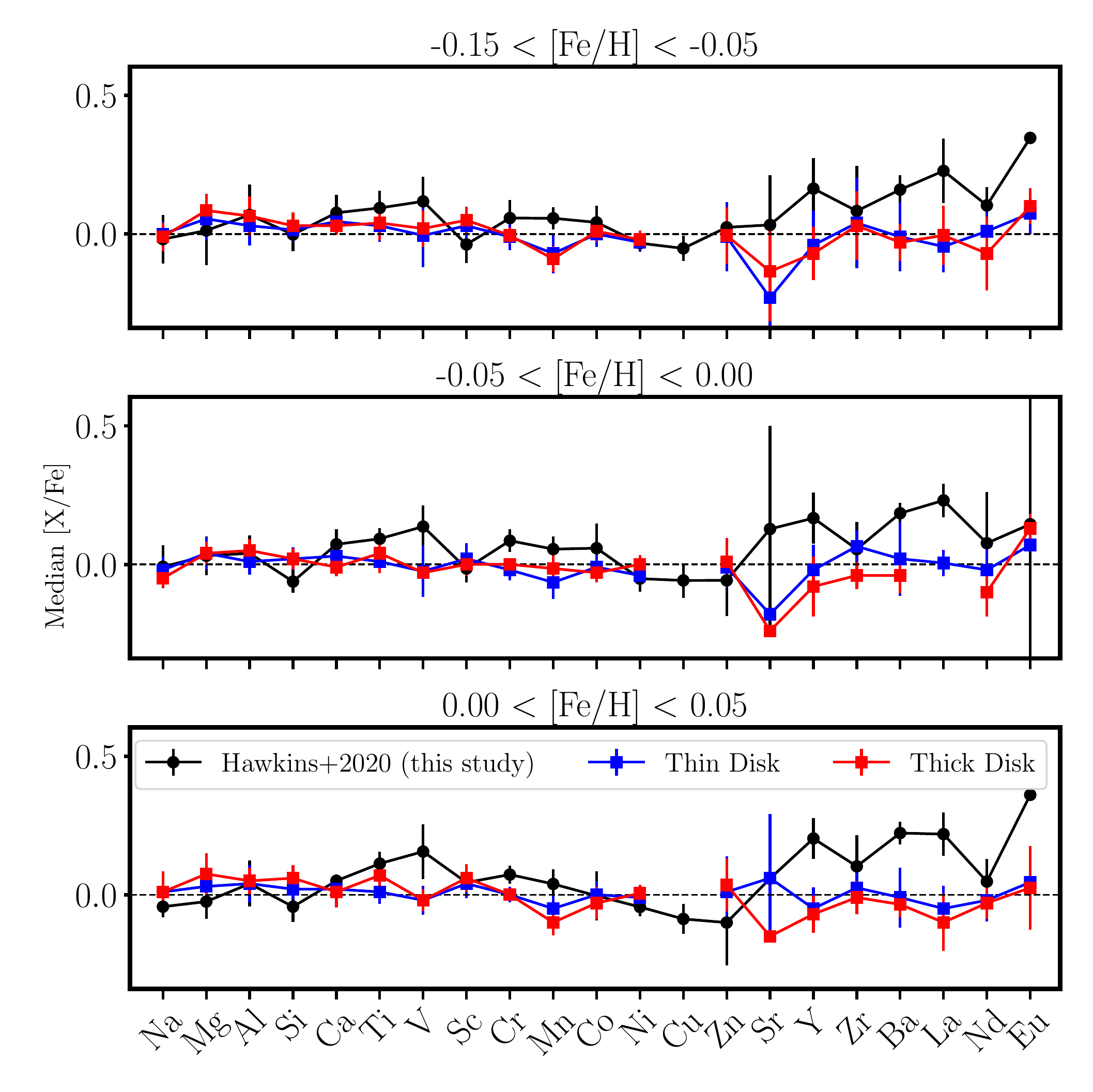}
	\caption{The median of the [X/Fe] abundance ratios for the \sname\ stream (black line). The errorbars represent the 1-$\sigma$ dispersion around the mean. For reference, the median (and dispersion) atmospheric [X/Fe] abundance ratios are shown for the Galactic thick (red line) and thin disk (blue line) in three metallicity bins corresponding to --0.15 $<$ \feh $<$ --0.05~dex (top panel),  --0.05 $<$ \feh $<$ 0.00~dex (middle panel), and 0.00 $<$ \feh $<$ 0.05~dex (bottom panel). The abundances for the Galactic thick and thin disks are sourced from \protect\citet[Na, Mg, Al, Si, Ti, Cr, Ni, Zn, Y, Ba, ][]{Bensby2014}, \protect\citet[Sc, V, Mn, Co,][]{Battistini2015}, and  \protect\citet[Sr, Zr, La, Ce, Nd, Eu,][]{Battistini2016}.  } 
	\label{fig:medXFE}
\end{figure}

In Fig.~\ref{fig:alpha} we show the [Mg, Si, Ca, Ti/Fe] abundance ratios as a function of \feh\ for the \sname\ stream (as black circles). For reference, we also show the behavior of the Galactic disk in all elemental abundance ratios \citep{Bensby2014, Adibekyan2012}. We find that the median [Mg, Si, Ca, Ti/Fe] are 0.01, --0.06, 0.07, 0.10~dex, respectively. These abundance ratios indicate that the \sname\ stream most resembles the Galactic thin disk, as expected. This is illustrated in Fig.~\ref{fig:medXFE}, which shows the median [X/Fe] for each element in \sname\ (black circles), relative to the Galactic thin disk (blue squares), and thick disk (red square). The median [X/Fe] for the thick and thin disk are determined from abundance measurements of 714 FGK disk dwarf stars in from \cite{Bensby2014}, \cite{Battistini2015} and \cite{ Battistini2016}. We remind the reader that there may be systematic offsets between the abundances derived in these works than those derived here due to different input physics (i.e. different atomic and molecular line lists, different line selections, etc.).  Stars are assigned to the thick and thin disks using their kinematics following the procedure outlined in \cite{Bensby2014}. We also find that the dispersion in these elements are all between 0.06--0.09~dex. However, since the typical uncertainty for these elements are generally less than 0.05~dex, we resolve a small abundance dispersion in the $\alpha$ elements. 

\begin{figure*}
	 \includegraphics[width=2\columnwidth]{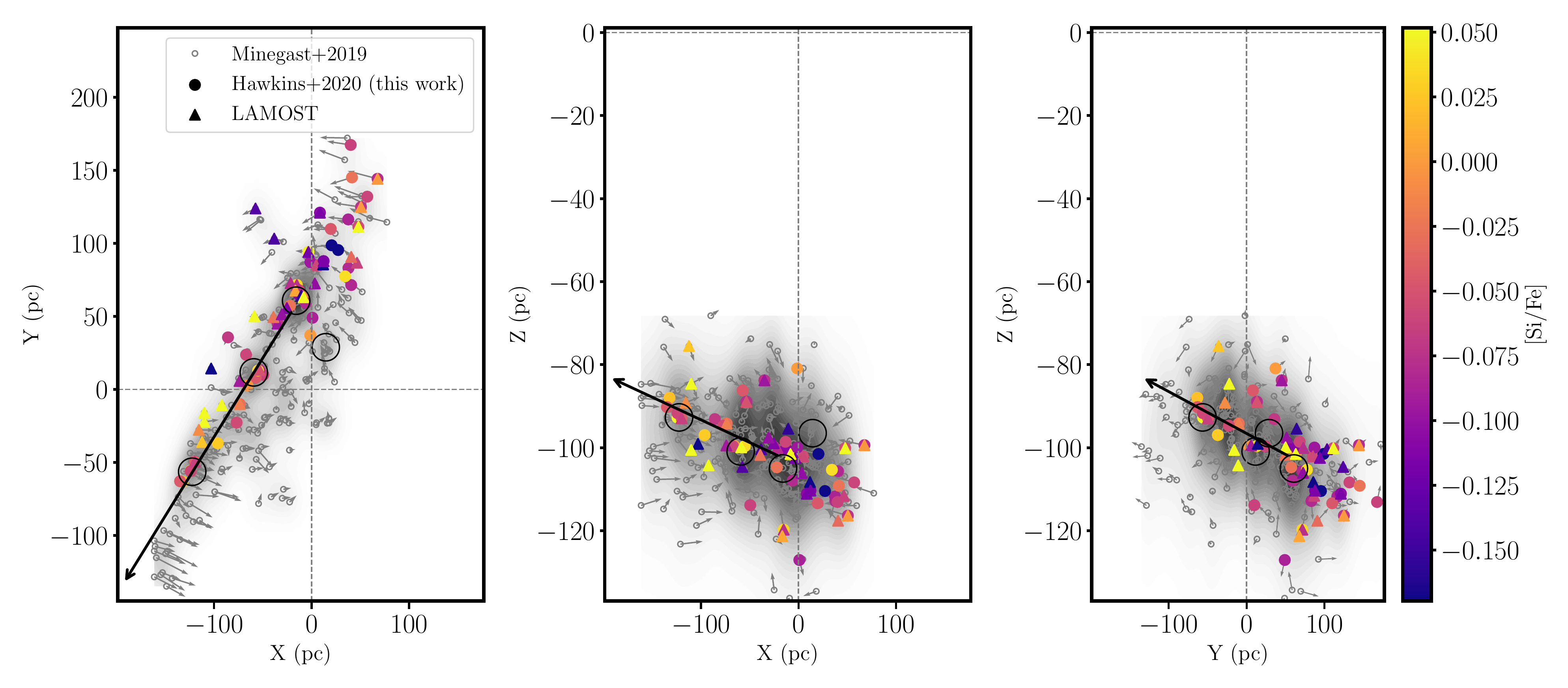}
	\caption{The Galactic X-Y (left panel), X-Z (middle panel), and Y-Z (right panel) coordinate projections of the \sname\ stream. For reference, the location of the Sun is at (0,0) in all plots. All stream members from \protect\cite{Meingast2019b} are shown as the open gray circles. The gray density in the background is a Gaussian Kernel density estimator the these data. Stars are color-coded by [Si/Fe], where known, either by LAMOST (triangles) or high-resolution spectra from this work  (circles). This illustrates a potentially weak gradient in [Si/Fe] along the major axis of the stream.  The large black arrows in each sub panel represent the major axis of the stream. However, unlike in Fig.~\ref{fig:met_across}, the smaller grey arrows are the velocity of the stars in \sname\ relative to its median systemic velocity. This enables us to view the internal velocity structure of the stream for the first time. The left panel makes it clear that the central part of \sname\ is moving at the systemic velocity while the outer edges of the cylindrical-shaped system is rotating around its `core'. One side of the outer edge of the cylinder is potentially slightly more enriched in [Si/Fe] compared to the other side. }
	\label{fig:si_across}
\end{figure*}

Similar to \feh\ (e.g. see Fig.~\ref{fig:met_across}), we also determined the gradient of [X/Fe] along the principal axis of the stream. Interestingly, we find that of all of the $\alpha$ elements, there is only a {\it weak} but significant (p$<$0.01) gradient in [Si/Fe] along the primary axis of the stream, with a correlation coefficient of 0.40 using data from our high-resolution dataset. This weak gradient persists whether we use our high-resolution data set or use the LAMOST dataset by itself (correlation coefficient of 0.32 with a p $\sim$ 0.05). It also persists even if we attempt to correct for any correlation between [Si/Fe] and \teff. To illustrate this gradient, in Fig.~\ref{fig:si_across}, we show how [Si/Fe] varies along the stream in Galactic X-Y (left panel), Y-Z (middle panel), X-Z spatial (right panel) projections. Similar to Fig.~\ref{fig:met_across}, we also show in the background all \sname\  candidates from \cite{Meingast2019b} and mark where there may be overdensities. However, unlike Fig.~\ref{fig:met_across}, in this figure we show the residuals of the 3-D velocity with respect to the median systemic velocity of the stream as small gray arrows in each sub panel in Fig.~\ref{fig:si_across}. This enables us to view the internal velocity distribution. We find that the `core' of the stream is moving at the systemic velocity of the stream, while the outer edges of the stream appear to rotate around the `core'.  Fig.~\ref{fig:si_across} indicates that the [Si/Fe] on one side of this outer shell may be more enriched than the other side. Further testing whether this gradient is astrophysical (or not) will undoubtedly require a significantly larger sample of \sname\ stars. This should be possible with the thousands of new candidate stream stars identified in \cite{Ratzenbock2020}. 

\subsection{odd-Z elements  (Na, Al, Sc, V, Cu)}
 \begin{figure}
	 \includegraphics[width=1\columnwidth]{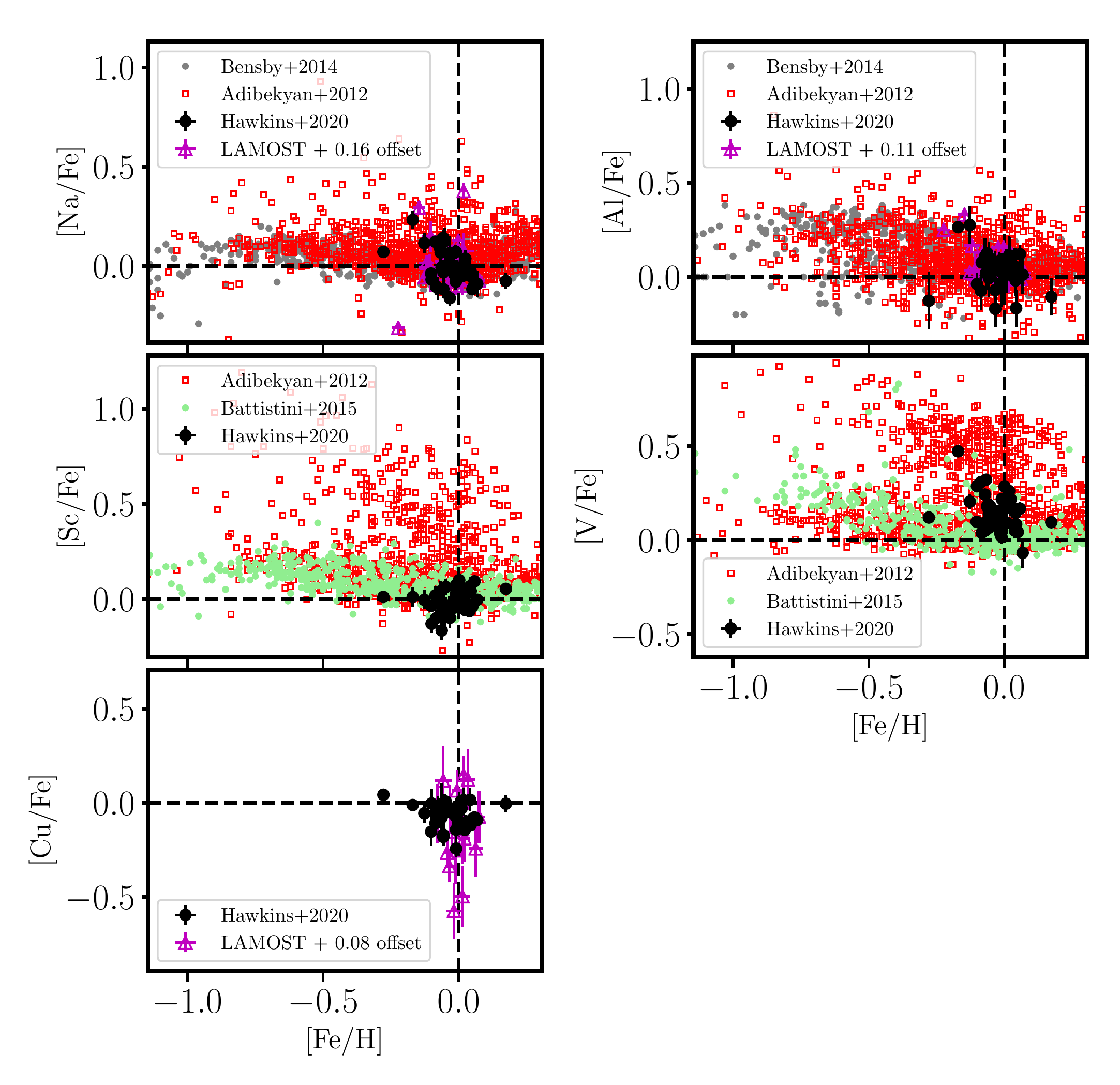}
	\caption{The atmospheric abundances ratios, [X/Fe], for the observed \sname\ stream candidates (black circles) for odd-Z (Na, Al, Sc, V, Cu)
	elements as a function of metallicity. For reference, behavior of each element with respect to metallicity is also shown for the Galactic disk. The reference data for the Galactic disk is taken from \protect\citet[gray circles,][]{Bensby2014}, \protect\citet[red open squares,][]{Adibekyan2012}, and \protect\citet[light green filled circles,][]{Battistini2015}.} 
	\label{fig:oddZ}
\end{figure}

The odd-Z elements studied here include Na, Al, Sc, V, and Cu. These elements are formed in a variety of ways. For example, Na is thought to be produced during hydrostatic C burning and during the NeNa cycle for massive stars. It is thought to be dispersed into the interstellar medium both through Type~II supernovae and partially through AGB stars. On the other hand, Al is thought to be produced during C and Ne burning as well as the MgAl cycle in massive stars. Al, much like Na, is thought to be dispersed into the interstellar medium via both Type~II supernovae and AGB stars \citep[e.g.][]{Samland1998, Kobayashi2006, Nomoto2013}. Sc and V are thought to be produced via explosive O and Ne burning, while Cu is largely produced via SNIa and hypernovae but is also produced by secondary phenomena in massive stars and the weak s-process \citep[e.g.][]{Samland1998, Mishenina2002, Andrievsky2018}. However, it is important to note that the nucleosynthesis pathway for the production of many of these elements are not well understood. This is indicated by the large discrepancy between the theoretical predictions of how these elements abundance ratio, [Sc, V, Cu/Fe] vary with \feh\ \citep[e.g.][and references therein]{Kobayashi2006, Nomoto2013}. 

In Fig.~\ref{fig:oddZ} we show the behavior of the odd-Z elements [Na, Al, Sc, V, Cu/Fe]  as a function of \feh\ for the \sname\ stream (as black circles). For reference, the behavior of the Galactic disk in all of the studied odd-Z elements with respect to metallicity are also displayed \citep{Bensby2014, Adibekyan2012, Battistini2015, Battistini2016}. We find that the median [Na, Al, Sc, V, Cu/ Fe] are --0.02, 0.04,  0.00, 0.14, 0.06~dex, respectively. These abundance ratios, similar to the $\alpha$ elements,  indicate that the \sname\ stream resembles the Galactic thin disk (Fig.~\ref{fig:medXFE}). The dispersion in these elements are all between 0.06--0.08~dex. However since the typical uncertainty for these elements are less than 0.06~dex, we resolve a very small abundance dispersion in the most odd-Z elements. We do not find any significant gradient of  [Na, Al, Sc, V, Cu/ Fe] along the principal axis of the stream.

\subsection{Fe-peak elements (Cr, Mn, Co, Ni, Zn)}
 \begin{figure}
	 \includegraphics[width=1\columnwidth]{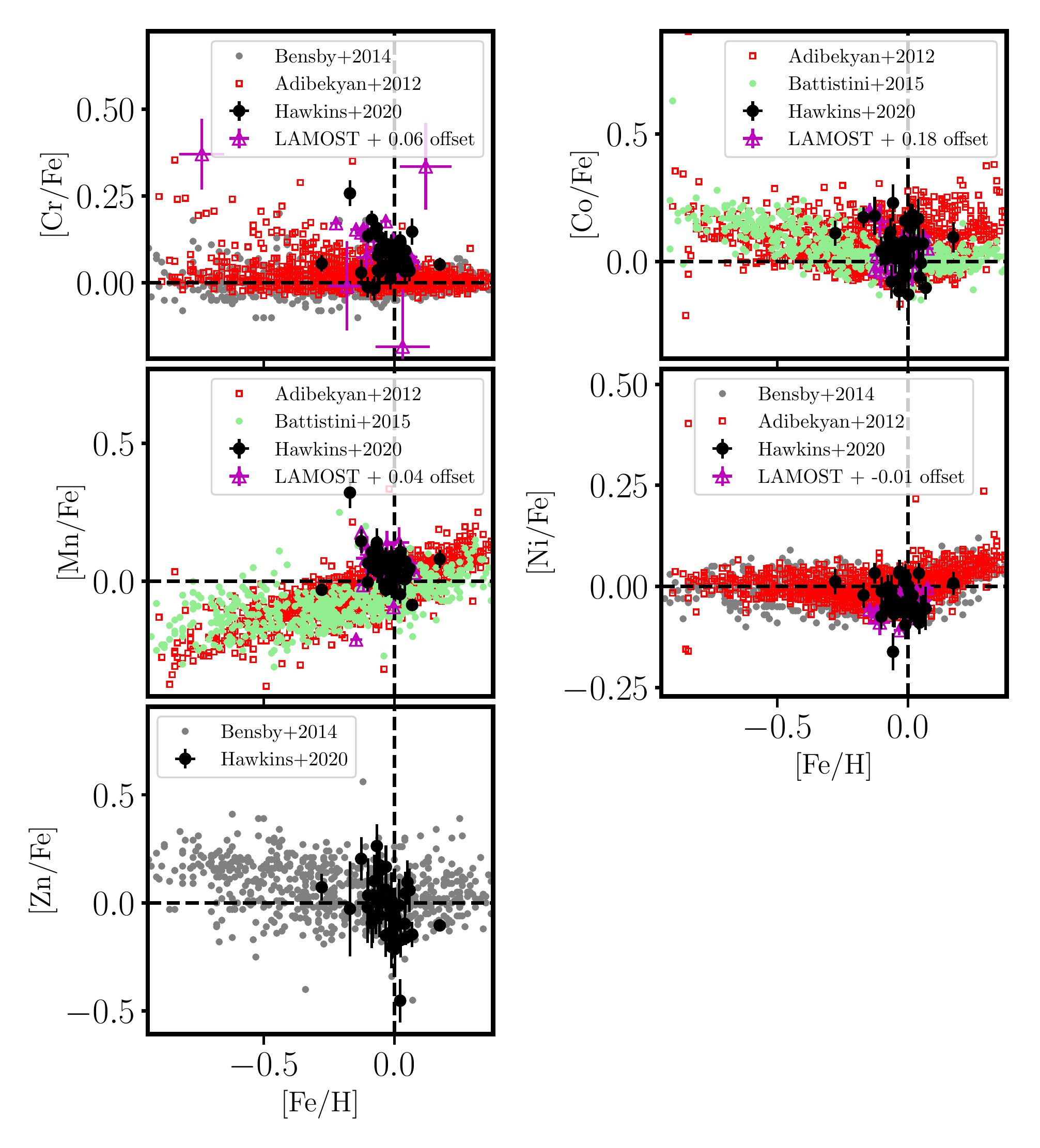}
	\caption{The atmospheric abundances ratios, [X/Fe], for the observed \sname\ stream candidates (black circles) Fe-peak (Cr, Mn, Co, Ni, Zn) elements as a function of metallicity. The symbols are the same as Fig.~\ref{fig:alpha} and Fig.\ref{fig:oddZ}.} 
	\label{fig:Fepeak}
\end{figure}
The Fe-peak elements, while formed in a variety of ways, are largely thought to be ejected into the interstellar medium through Type~Ia and Type~II supernovae and hypernovae \citep[e.g.][]{Samland1998, Iwamoto1999, Kobayashi2006, Kobayashi2011, Nomoto2013}. For example, Mn, Ni, and Cr are thought to be produced by explosive Si burning in supernovae while Co and Zn are thought to be produced in hypernovea \citep[e.g. see ][for a more detailed discussion on the formation of the Fe-peak elements]{Samland1998, Kobayashi2006, Nomoto2013, Mikolaitis2017}. While these elements were initially thought to largely track Fe, there are some differences. For example, Mn is produced at significantly higher levels in Type~Ia supernovae and thus [Mn/Fe] shows a decreasing trend with increasing \feh\ \citep{Gratton1989, Kobayashi2011}. 

In Fig.~\ref{fig:Fepeak}, we show the behavior of the Fe-peak elements [Cr, Mn, Co, Ni, Zn/Fe]  as a function of \feh\ for the \sname\ stream (as black circles). As before, we also show the behavior of the Galactic disk in this figure. We find that the median [Cr, Mn, Co, Ni, Zn/Fe] are 0.08, 0.05, 0.05, --0.03, --0.04~dex, respectively. These abundance ratios indicate that the \sname\ stream most resembles the Galactic thin disk (also see Fig.~\ref{fig:medXFE}). The dispersion in these elements are all between 0.06--0.13~dex. We just (at the $\sim 1\sigma$ level) resolve significant abundance dispersion in the Fe-peak elements. Similarly to the odd-Z elements, we do not find any significant gradient of [Cr, Mn, Co, Ni, Zn/ Fe] along the principal axis of the stream.

\subsection{Neutron Capture  elements (Sr, Y, Zr, Ba, La, Nd, Eu)}
\begin{figure}
	 \includegraphics[width=1\columnwidth]{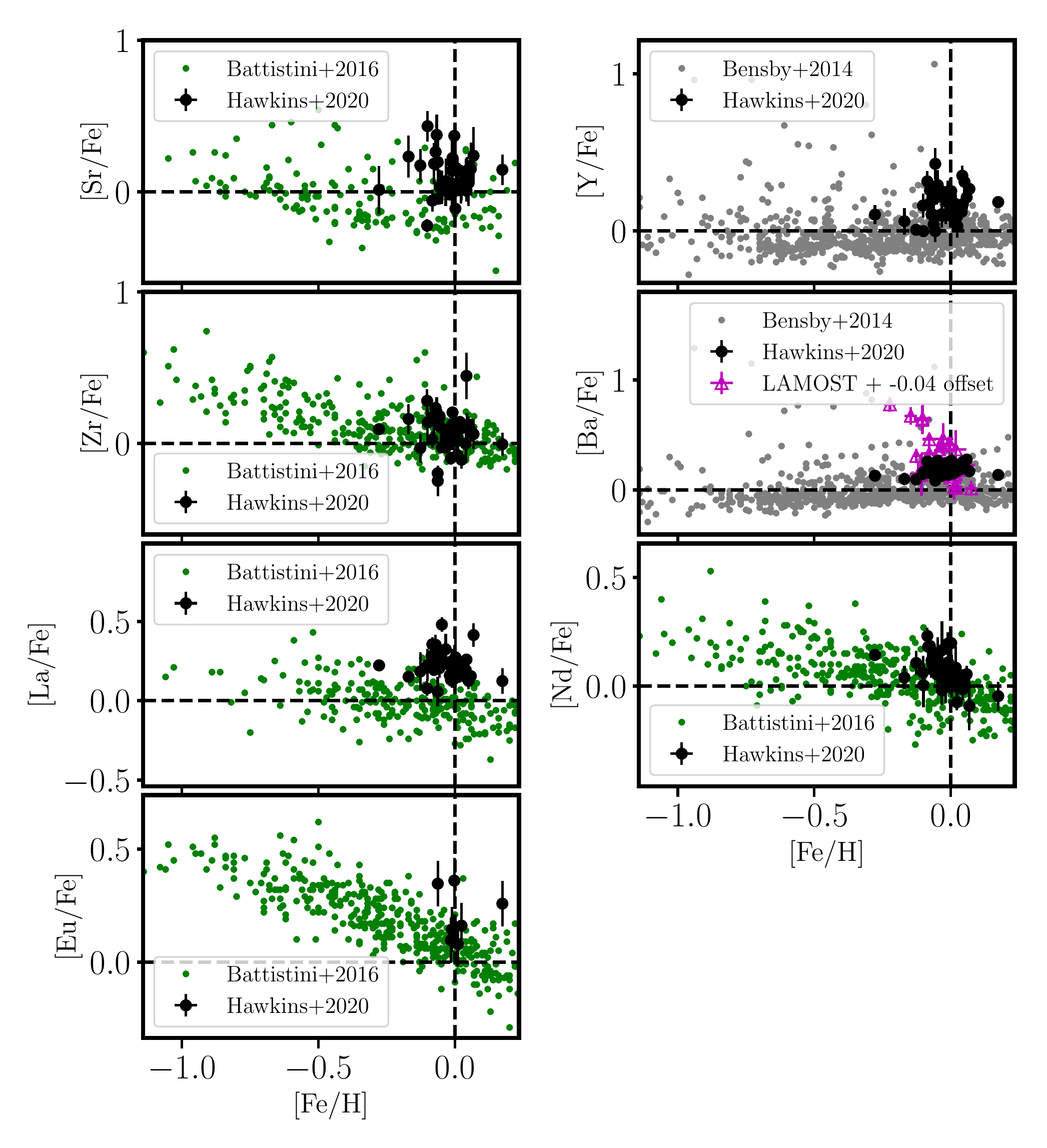}
	\caption{The atmospheric abundances ratios, [X/Fe], for the observed \sname\ stream candidates (black circles) for the neutron capture elements (Sr, Zr, La, Eu, Y, Ba, Nd) as a function of metallicity. For reference, also shown is the behavior of each element relative to metallically is shown from    \protect\citet[light green circles,][]{Bensby2014}, \protect\citet[dark green circles,][]{Battistini2016}. } 
	\label{fig:ncap}
\end{figure}

The neutron capture elements are thought to form in variety ways. As such, they are often separated in two groups which depend on the formation channel. The first of these is the slow neutron capture (s-process) elements, which include primarily Sr, Y, Zr, Ba, La, and Nd. These are thought to form through the slow neutron capture process in AGB stars and are then subsequently ejected through stellar winds. On the other hand the rapid neutron capture process (r-process), which requires a significantly higher neutron flux compared to the s-process, is thought to create much of the Eu in the universe. The r-process site is still under debate but binary neutron stars are currently the favored mechanism \citep[e.g][]{Vandevoort2019}. 
In Fig.~\ref{fig:ncap} we show the behavior of the Fe-peak elements [Sr, Y, Zr, Ba, La, Nd, Eu/Fe]  as a function of \feh\ for the \sname\ stream (as black circles). We find that the median [Sr, Y, Zr, Ba, La, Nd, Eu/Fe] are 0.12, 0.18, 0.08, 0.19, 0.22, 0.07, 0.21~dex, respectively. For reference, we also how these elemental ratios vary as a function of \feh\ in the Galactic disk \citep{Bensby2014, Battistini2015}. We find that the \sname\ stream is somewhat enhanced in the neutron capture elements Sr, Y,  Ba, La, and Eu relative to the Galactic disk (also see Fig.~\ref{fig:medXFE}). One reason for this could be due to systematic trends between these elemental ratios and \teff, which are not usually expected in the cluster regime, enrichment (self or othewise) or some other unknown reason. Larger samples with higher-quality (specifically higher SNR) data will be required to further understand the level of any enhancement.  

The dispersions in these elements are significantly larger than the other studied elements and range between 0.08--0.28~dex. However, we find that the largest dispersion exist in elements where there is typically only one usable line (e.g. Eu, Sr). Nonetheless, we generally find that the the dispersion in the [X/Fe] abundance ratio for the neutron capture elements are larger, by a factor of $\sim$2, compared to the typical uncertainty. Similar to the odd-Z elements, we do not find any significant gradient of  [Sr, Y, Zr, Ba, La, Nd, Eu/ Fe] along the principal axis of the stream.

\section{Discussion : Formation, Age, and Chemistry of the Psc--Eri Stream}\label{sec:discussion}

The \sname\ stream of stars flowing through the Galactic disk represents one the most massive coeval populations of its type in the Solar neighborhood. When it was first discovered by \cite{Meingast2019b}, it was suggested that its age was consistent with $\sim$1~Gyr. This was based on isochrone fitting of the color--magnitude diagram of the 256 members found, anchored by the giant system 42 Cet. Follow-up work in gyrochronology \citep{Curtis2019} and color--magnitude diagram isochrone fitting the expanded list of $>$1000 members \citep{Roser2020, Ratzenbock2020}, have indicated that the stream is only $\approx$120~Myr old. In this work, we have shown that the behaviour of Li with respect to \teff\ (see Fig.~\ref{fig:Li}) rules out the older age of the stream and implies that it is only $\sim$120--200 Myr. This is consistent with a recent result on the Li equivalent widths of 40 GK-type stream members by \cite{Arancibia-Silva2020}. 

The \sname\ stream is thought to be a coherent, coeval young aggregation of stars moving together with a common origin (i.e., formed from a single molecular cloud). This is expected as most stars are thought to form in clustered, possibly spherical-like, configuration \citep[e.g.][and references therein]{Lada2003}. However, stellar feedback from massive stars forming in the stream can disrupt the molecular cloud that it is forming from, causing it to become stream- or filament-like, or potentially even completely unbound from the initial stellar distribution \citep[e.g.][]{Tutukov1978}. There have been recent examples of filamentary-like relics of recent ($<$ 100~Myr) star formation sites \citep[e.g. in the Vela OB2 and Orion star-forming regions, ][]{Beccari2020, Jerabkova2019}. These examples may suggest that the \sname\ stream may in fact be a relic of star formation along a filament.   Regardless, one might expect the stream to be chemically homogeneous as it is formed from a common molecular cloud. However, if that molecular cloud that formed the stream was sufficiently large there may be chemical differences across the cloud which could propagate into inhomogeneity in the \sname\ stream \citep[e.g.][]{Krumholz2018}. This is particularly true in the neutron capture elements. While we observe some dispersion in these elements, their relative uncertainties are also larger and therefore this is worth a larger exploration with a larger sample. 

\sname\ is similar to the Pleiades in age and mass, but is strikingly different in its morphology in that the \sname\ stream is a long cylindrical cigar-shaped object spanning 700~pc in space, while the Pleiades is a denser, more spherically-shaped self-gravitating open cluster. We find that the \sname\ stream has a noticeable, but not large, \feh\ dispersion (of 0.07~dex) and dispersions of similar size in other elements. We also find that the measured abundance dispersions within the \sname\ cluster are larger in the neutron capture elements. The elongated structure of the \sname\ stream, along with chemical inhomogeneity in the giant molecular cloud(s) that it formed from could explain why the \sname\ stream is observed to have noticeable, but not large, chemical abundance dispersions especially in the neutron capture elements \citep{Krumholz2018}.

\section{Summary}\label{sec:conclusions}
The \sname\ stream is a newly discovered coeval population discovered by \cite{Meingast2019b}. It is was initially proposed to be 1~Gyr old, containing up to $\sim$2000~\Msun\ \citep{Meingast2019b}. More recent studies of the stream, however, note that it may be a significantly younger stream through fitting isochrones to the upper main sequence that was missing from the original \citet{Meingast2019b} list \citep{Curtis2019, Roser2020, Ratzenbock2020}, gyrochronology \citep{Curtis2019}, and the Li depletion pattern \citep{Arancibia-Silva2020}. In this work, we explore the {\it detailed } chemistry of \numstar\ \sname\ stream stars with high-resolution (R$\sim$ 60000) optical spectroscopic data from McDonald Observatory as well as 40 stream stars from low-resolution (R$\sim$1800) LAMOST data.  

Our results indicate the following regarding the \sname\ stream:
\begin{enumerate}
    \item It has a metallicity distribution that is peaked at just below Solar with \feh\ =  --0.03~dex. Additionally it has a rather low, but measurable, dispersion in metallicity, with $\sigma$\feh\ = 0.07~dex. The dispersion is reduced to $\sigma$\feh\ = 0.04~dex if we restrict the sample to a high-probability set of  \sname\ members that have been identified by both \cite{Meingast2019b} and \cite{Roser2020}. A larger sample of high-resolution data is required to further determine the homogeneity of the cluster.
    
    \item The stream appears to have abundance dispersions that are only very slightly larger than the typical 1$\sigma$ uncertainties for many elements including \feh.

    \item Psc--Eri and the Pleiades are approximately coeval ($\approx$120~Myr), according to their \teff--A(Li) distributions (see Fig.~\ref{fig:Li}). 
    
    \item The stream shows no significant abundance gradients along its major axis, except in [Si/Fe]. This gradient of [Si/Fe] persists in both the LAMOST data and our high-resolution data as well. Larger studies of the detailed chemistry of this stream are needed to confirm (or refute) the existence of this gradient. 
    
    \item The detailed chemistry of the stream matches the expectations of the local Galactic thin disk; however, some enhancements are seen in several neutron capture element (e.g., in Sr, Y, Ba, and La). These enhancements could be a result of enrichment or due to an unknown systematic.

    \item  The stream morphologically appears to be a cylindrical-like system whose outer parts are rotating about the center (e.g. Fig.~\ref{fig:si_across}).
    
\end{enumerate}
These results, taken together, rule out the initial old age estimate for the Psc--Eri. They also illustrate the power of studying the detailed chemistry (from Li to Eu) as a way to further constrain the age and formation of the stream. These results show that this newly discovered  cylindrical-shaped stream is both young \citep[according to the Li studied here and in][]{Arancibia-Silva2020}, and that it is also a chemically interesting laboratory for studying star and planet formation physics. A detailed chemical analysis of a significantly larger sample of \sname\ stars will be required to not only confirm (or not) the abundance dispersion and the [Si/Fe] gradient but also to determine whether the overdensities along the stream are unique or formed from well-mixed gas on the several hundred parsec scale.

\section*{Acknowledgements}
We thank Charlie Conroy for useful comments on this work. This paper includes data taken at The McDonald Observatory of The University of Texas at Austin. We thank the staff at McDonald Observatory for making this project possible. KH has been partially supported by a TDA/Scialog  (2018-2020) grant funded by the Research Corporation and a TDA/Scialog grant (2019-2021) funded by the Heising-Simons Foundation. KH acknowledges support from the National Science Foundation grant AST-1907417. KH and  ML acknowledge support from the Wootton Center for Astrophysical Plasma Properties under the United States Department of Energy collaborative agreement DE-NA0003843.

This work has made use of data from the European Space Agency (ESA)
mission {\it Gaia} (\url{https://www.cosmos.esa.int/gaia}), processed by
the {\it Gaia} Data Processing and Analysis Consortium (DPAC,
\url{https://www.cosmos.esa.int/web/gaia/dpac/consortium}). Funding
for the DPAC has been provided by national institutions, in particular
the institutions participating in the {\it Gaia} Multilateral Agreement.

Guoshoujing Telescope (the Large Sky Area Multi-Object Fiber Spectroscopic Telescope LAMOST) is a National Major Scientific Project built by the Chinese Academy of Sciences. Funding for the project has been provided by the National Development and Reform Commission. LAMOST is operated and managed by the National Astronomical Observatories, Chinese Academy of Sciences.



\bibliographystyle{mnras}
\bibliography{bibliography.bib} 

\begin{thebibliography}{}
\makeatletter
\relax
\def\mn@urlcharsother{\let\do\@makeother \do\$\do\&\do\#\do\^\do\_\do\%\do\~}
\def\mn@doi{\begingroup\mn@urlcharsother \@ifnextchar [ {\mn@doi@}
  {\mn@doi@[]}}
\def\mn@doi@[#1]#2{\def\@tempa{#1}\ifx\@tempa\@empty \href
  {http://dx.doi.org/#2} {doi:#2}\else \href {http://dx.doi.org/#2} {#1}\fi
  \endgroup}
\def\mn@eprint#1#2{\mn@eprint@#1:#2::\@nil}
\def\mn@eprint@arXiv#1{\href {http://arxiv.org/abs/#1} {{\tt arXiv:#1}}}
\def\mn@eprint@dblp#1{\href {http://dblp.uni-trier.de/rec/bibtex/#1.xml}
  {dblp:#1}}
\def\mn@eprint@#1:#2:#3:#4\@nil{\def\@tempa {#1}\def\@tempb {#2}\def\@tempc
  {#3}\ifx \@tempc \@empty \let \@tempc \@tempb \let \@tempb \@tempa \fi \ifx
  \@tempb \@empty \def\@tempb {arXiv}\fi \@ifundefined
  {mn@eprint@\@tempb}{\@tempb:\@tempc}{\expandafter \expandafter \csname
  mn@eprint@\@tempb\endcsname \expandafter{\@tempc}}}

\bibitem[\protect\citeauthoryear{{Adibekyan}, {Sousa}, {Santos}, {Delgado
  Mena}, {Gonz{\'a}lez Hern{\'a}ndez}, {Israelian}, {Mayor}  \&
  {Khachatryan}}{{Adibekyan} et~al.}{2012}]{Adibekyan2012}
{Adibekyan} V.~Z.,  {Sousa} S.~G.,  {Santos} N.~C.,  {Delgado Mena} E.,
  {Gonz{\'a}lez Hern{\'a}ndez} J.~I.,  {Israelian} G.,  {Mayor} M.,
  {Khachatryan} G.,  2012, \mn@doi [\aap] {10.1051/0004-6361/201219401}, \href
  {http://adsabs.harvard.edu/abs/2012A%26A...545A..32A} {545, A32}

\bibitem[\protect\citeauthoryear{{Alvarez} \& {Plez}}{{Alvarez} \&
  {Plez}}{1998}]{Alvarez1998}
{Alvarez} R.,  {Plez} B.,  1998, \aap, \href
  {http://cdsads.u-strasbg.fr/abs/1998A%26A...330.1109A} {330, 1109}

\bibitem[\protect\citeauthoryear{{Andrievsky}, {Bonifacio}, {Caffau},
  {Korotin}, {Spite}, {Spite}, {Sbordone}  \& {Zhukova}}{{Andrievsky}
  et~al.}{2018}]{Andrievsky2018}
{Andrievsky} S.,  {Bonifacio} P.,  {Caffau} E.,  {Korotin} S.,  {Spite} M.,
  {Spite} F.,  {Sbordone} L.,   {Zhukova} A.~V.,  2018, \mn@doi [\mnras]
  {10.1093/mnras/stx2526}, \href
  {http://adsabs.harvard.edu/abs/2018MNRAS.473.3377A} {473, 3377}

\bibitem[\protect\citeauthoryear{{Arancibia-Silva}, {Bouvier}, {Bayo}, {Galli},
  {Brandner}, {Bouy}  \& {Barrado}}{{Arancibia-Silva}
  et~al.}{2020}]{Arancibia-Silva2020}
{Arancibia-Silva} J.,  {Bouvier} J.,  {Bayo} A.,  {Galli} P.~A.~B.,  {Brandner}
  W.,  {Bouy} H.,   {Barrado} D.,  2020, arXiv e-prints, \href
  {https://ui.adsabs.harvard.edu/abs/2020arXiv200210556A} {p. arXiv:2002.10556}

\bibitem[\protect\citeauthoryear{{Asplund}, {Nordlund}, {Trampedach}  \&
  {Stein}}{{Asplund} et~al.}{2000}]{Asplund2000b}
{Asplund} M.,  {Nordlund} {\r{A}}.,  {Trampedach} R.,   {Stein} R.~F.,  2000,
  \aap, \href {https://ui.adsabs.harvard.edu/abs/2000A&A...359..743A} {359,
  743}

\bibitem[\protect\citeauthoryear{{Asplund}, {Grevesse}  \& {Sauval}}{{Asplund}
  et~al.}{2005}]{Asplund2005}
{Asplund} M.,  {Grevesse} N.,   {Sauval} A.~J.,  2005, in {Barnes} III T.~G.,
  {Bash} F.~N.,  eds,  Astronomical Society of the Pacific Conference Series
  Vol. 336, Cosmic Abundances as Records of Stellar Evolution and
  Nucleosynthesis. p.~25

\bibitem[\protect\citeauthoryear{{Battistini} \& {Bensby}}{{Battistini} \&
  {Bensby}}{2015}]{Battistini2015}
{Battistini} C.,  {Bensby} T.,  2015, \mn@doi [\aap]
  {10.1051/0004-6361/201425327}, \href
  {http://adsabs.harvard.edu/abs/2015A%26A...577A...9B} {577, A9}

\bibitem[\protect\citeauthoryear{{Battistini} \& {Bensby}}{{Battistini} \&
  {Bensby}}{2016}]{Battistini2016}
{Battistini} C.,  {Bensby} T.,  2016, \mn@doi [\aap]
  {10.1051/0004-6361/201527385}, \href
  {http://adsabs.harvard.edu/abs/2016A%26A...586A..49B} {586, A49}

\bibitem[\protect\citeauthoryear{{Beccari}, {Boffin}  \& {Jerabkova}}{{Beccari}
  et~al.}{2020}]{Beccari2020}
{Beccari} G.,  {Boffin} H. M.~J.,   {Jerabkova} T.,  2020, \mn@doi [\mnras]
  {10.1093/mnras/stz3195}, \href
  {https://ui.adsabs.harvard.edu/abs/2020MNRAS.491.2205B} {491, 2205}

\bibitem[\protect\citeauthoryear{{Bensby}, {Feltzing}  \& {Oey}}{{Bensby}
  et~al.}{2014}]{Bensby2014}
{Bensby} T.,  {Feltzing} S.,   {Oey} M.~S.,  2014, \mn@doi [\aap]
  {10.1051/0004-6361/201322631}, \href
  {http://adsabs.harvard.edu/abs/2014A%26A...562A..71B} {562, A71}

\bibitem[\protect\citeauthoryear{{Blanco-Cuaresma}, {Soubiran}, {Heiter}  \&
  {Jofr{\'e}}}{{Blanco-Cuaresma} et~al.}{2014}]{2014A&A...569A.111B}
{Blanco-Cuaresma} S.,  {Soubiran} C.,  {Heiter} U.,   {Jofr{\'e}} P.,  2014,
  \mn@doi [\aap] {10.1051/0004-6361/201423945}, \href
  {http://adsabs.harvard.edu/abs/2014A%26A...569A.111B} {569, A111}

\bibitem[\protect\citeauthoryear{{Boeche}, {Smith}, {Grebel}, {Zhong}, {Hou},
  {Chen}  \& {Stello}}{{Boeche} et~al.}{2018}]{Boeche2018}
{Boeche} C.,  {Smith} M.~C.,  {Grebel} E.~K.,  {Zhong} J.,  {Hou} J.~L.,
  {Chen} L.,   {Stello} D.,  2018, \mn@doi [\aj] {10.3847/1538-3881/aab5af},
  \href {https://ui.adsabs.harvard.edu/abs/2018AJ....155..181B} {155, 181}

\bibitem[\protect\citeauthoryear{{Boesgaard}, {Deliyannis}, {Stephens}  \&
  {King}}{{Boesgaard} et~al.}{1998}]{Boesgaard1998}
{Boesgaard} A.~M.,  {Deliyannis} C.~P.,  {Stephens} A.,   {King} J.~R.,  1998,
  \mn@doi [\apj] {10.1086/305089}, \href
  {https://ui.adsabs.harvard.edu/abs/1998ApJ...493..206B} {493, 206}

\bibitem[\protect\citeauthoryear{{Bouvier} et~al.,}{{Bouvier}
  et~al.}{2018}]{Bouvier2018}
{Bouvier} J.,  et~al., 2018, \mn@doi [\aap] {10.1051/0004-6361/201731881},
  \href {https://ui.adsabs.harvard.edu/abs/2018A&A...613A..63B} {613, A63}

\bibitem[\protect\citeauthoryear{{Bovy}}{{Bovy}}{2016}]{Bovy2016}
{Bovy} J.,  2016, \mn@doi [\apj] {10.3847/0004-637X/817/1/49}, \href
  {http://adsabs.harvard.edu/abs/2016ApJ...817...49B} {817, 49}

\bibitem[\protect\citeauthoryear{{Buder} et~al.,}{{Buder}
  et~al.}{2018}]{Buder2018}
{Buder} S.,  et~al., 2018, preprint, \href
  {http://adsabs.harvard.edu/abs/2018arXiv180406041B} {} (\mn@eprint {arXiv}
  {1804.06041})

\bibitem[\protect\citeauthoryear{{Cargile}, {James}  \& {Jeffries}}{{Cargile}
  et~al.}{2010}]{Cargile2010}
{Cargile} P.~A.,  {James} D.~J.,   {Jeffries} R.~D.,  2010, \mn@doi [\apjl]
  {10.1088/2041-8205/725/2/L111}, \href
  {https://ui.adsabs.harvard.edu/abs/2010ApJ...725L.111C} {725, L111}

\bibitem[\protect\citeauthoryear{{Carlos} et~al.,}{{Carlos}
  et~al.}{2019}]{Carlos2019}
{Carlos} M.,  et~al., 2019, \mn@doi [\mnras] {10.1093/mnras/stz681}, \href
  {https://ui.adsabs.harvard.edu/abs/2019MNRAS.485.4052C} {485, 4052}

\bibitem[\protect\citeauthoryear{{Charbonnel} \& {Talon}}{{Charbonnel} \&
  {Talon}}{2005}]{Charbonnel2005}
{Charbonnel} C.,  {Talon} S.,  2005, \mn@doi [Science]
  {10.1126/science.1116849}, \href
  {https://ui.adsabs.harvard.edu/abs/2005Sci...309.2189C} {309, 2189}

\bibitem[\protect\citeauthoryear{{Cui} et~al.,}{{Cui} et~al.}{2012}]{Cui2012}
{Cui} X.-Q.,  et~al., 2012, \mn@doi [Research in Astronomy and Astrophysics]
  {10.1088/1674-4527/12/9/003}, \href
  {https://ui.adsabs.harvard.edu/abs/2012RAA....12.1197C} {12, 1197}

\bibitem[\protect\citeauthoryear{{Cummings}, {Deliyannis}, {Anthony-Twarog},
  {Twarog}  \& {Maderak}}{{Cummings} et~al.}{2012}]{Cummings2012}
{Cummings} J.~D.,  {Deliyannis} C.~P.,  {Anthony-Twarog} B.,  {Twarog} B.,
  {Maderak} R.~M.,  2012, \mn@doi [\aj] {10.1088/0004-6256/144/5/137}, \href
  {https://ui.adsabs.harvard.edu/abs/2012AJ....144..137C} {144, 137}

\bibitem[\protect\citeauthoryear{{Curtis}, {Ag{\"u}eros}, {Mamajek}, {Wright}
  \& {Cummings}}{{Curtis} et~al.}{2019a}]{Curtis2019}
{Curtis} J.~L.,  {Ag{\"u}eros} M.~A.,  {Mamajek} E.~E.,  {Wright} J.~T.,
  {Cummings} J.~D.,  2019a, \mn@doi [\aj] {10.3847/1538-3881/ab2899}, \href
  {https://ui.adsabs.harvard.edu/abs/2019AJ....158...77C} {158, 77}

\bibitem[\protect\citeauthoryear{{Curtis}, {Ebinger}, {Fr{\"o}hlich}, {Hempel},
  {Perego}, {Liebend{\"o}rfer}  \& {Thielemann}}{{Curtis}
  et~al.}{2019b}]{Curtis2019sn}
{Curtis} S.,  {Ebinger} K.,  {Fr{\"o}hlich} C.,  {Hempel} M.,  {Perego} A.,
  {Liebend{\"o}rfer} M.,   {Thielemann} F.-K.,  2019b, \mn@doi [\apj]
  {10.3847/1538-4357/aae7d2}, \href
  {https://ui.adsabs.harvard.edu/abs/2019ApJ...870....2C} {870, 2}

\bibitem[\protect\citeauthoryear{{Davis}, {Cisewski}, {Dumusque}, {Fischer}  \&
  {Ford}}{{Davis} et~al.}{2017}]{Davis2017}
{Davis} A.~B.,  {Cisewski} J.,  {Dumusque} X.,  {Fischer} D.~A.,   {Ford}
  E.~B.,  2017, \mn@doi [\apj] {10.3847/1538-4357/aa8303}, \href
  {https://ui.adsabs.harvard.edu/abs/2017ApJ...846...59D} {846, 59}

\bibitem[\protect\citeauthoryear{{De Silva} et~al.,}{{De Silva}
  et~al.}{2015}]{De_silva2015}
{De Silva} G.~M.,  et~al., 2015, \mn@doi [\mnras] {10.1093/mnras/stv327}, \href
  {http://adsabs.harvard.edu/abs/2015MNRAS.449.2604D} {449, 2604}

\bibitem[\protect\citeauthoryear{{Dias}, {L{\'e}pine}  \& {Alessi}}{{Dias}
  et~al.}{2002}]{Dias2002}
{Dias} W.~S.,  {L{\'e}pine} J.~R.~D.,   {Alessi} B.~S.,  2002, \mn@doi [\aap]
  {10.1051/0004-6361:20020417}, \href
  {https://ui.adsabs.harvard.edu/abs/2002A&A...388..168D} {388, 168}

\bibitem[\protect\citeauthoryear{{Dumusque}}{{Dumusque}}{2018}]{Dumusque2018}
{Dumusque} X.,  2018, \mn@doi [\aap] {10.1051/0004-6361/201833795}, \href
  {https://ui.adsabs.harvard.edu/abs/2018A&A...620A..47D} {620, A47}

\bibitem[\protect\citeauthoryear{{Gaia Collaboration}, {Brown}, {Vallenari},
  {Prusti}, {de Bruijne}, {Babusiaux}  \& {Bailer-Jones}}{{Gaia Collaboration}
  et~al.}{2018}]{Gaiasummary2018}
{Gaia Collaboration} {Brown} A.~G.~A.,  {Vallenari} A.,  {Prusti} T.,  {de
  Bruijne} J.~H.~J.,  {Babusiaux} C.,   {Bailer-Jones} C.~A.~L.,  2018,
  preprint, \href {http://adsabs.harvard.edu/abs/2018arXiv180409365G} {}
  (\mn@eprint {arXiv} {1804.09365})

\bibitem[\protect\citeauthoryear{{Gratton}}{{Gratton}}{1989}]{Gratton1989}
{Gratton} R.~G.,  1989, \aap, \href
  {http://adsabs.harvard.edu/abs/1989A%26A...208..171G} {208, 171}

\bibitem[\protect\citeauthoryear{{Gustafsson}, {Edvardsson}, {Eriksson},
  {J{\o}rgensen}, {Nordlund}  \& {Plez}}{{Gustafsson}
  et~al.}{2008}]{Gustafsson2008}
{Gustafsson} B.,  {Edvardsson} B.,  {Eriksson} K.,  {J{\o}rgensen} U.~G.,
  {Nordlund} {\AA}.,   {Plez} B.,  2008, \mn@doi [\aap]
  {10.1051/0004-6361:200809724}, \href
  {http://adsabs.harvard.edu/abs/2008A%26A...486..951G} {486, 951}

\bibitem[\protect\citeauthoryear{{Hawkins}, {Jofr{\'e}}, {Masseron}  \&
  {Gilmore}}{{Hawkins} et~al.}{2015}]{Hawkins2015b}
{Hawkins} K.,  {Jofr{\'e}} P.,  {Masseron} T.,   {Gilmore} G.,  2015, \mn@doi
  [\mnras] {10.1093/mnras/stv1586}, \href
  {http://adsabs.harvard.edu/abs/2015MNRAS.453..758H} {453, 758}

\bibitem[\protect\citeauthoryear{{Hawkins} et~al.,}{{Hawkins}
  et~al.}{2020}]{Hawkins2019}
{Hawkins} K.,  et~al., 2020, \mn@doi [\mnras] {10.1093/mnras/stz3132}, \href
  {https://ui.adsabs.harvard.edu/abs/2020MNRAS.492.1164H} {492, 1164}

\bibitem[\protect\citeauthoryear{{Iwamoto}, {Brachwitz}, {Nomoto}, {Kishimoto},
  {Umeda}, {Hix}  \& {Thielemann}}{{Iwamoto} et~al.}{1999}]{Iwamoto1999}
{Iwamoto} K.,  {Brachwitz} F.,  {Nomoto} K.,  {Kishimoto} N.,  {Umeda} H.,
  {Hix} W.~R.,   {Thielemann} F.-K.,  1999, \mn@doi [\apjs] {10.1086/313278},
  \href {http://adsabs.harvard.edu/abs/1999ApJS..125..439I} {125, 439}

\bibitem[\protect\citeauthoryear{{Jeffries}}{{Jeffries}}{2000}]{Jeffries2000}
{Jeffries} R.~D.,  2000, {Lithium depletion in open clusters}.
Astronomical Society of the Pacific, p.~245

\bibitem[\protect\citeauthoryear{{Jeffries}}{{Jeffries}}{2014}]{Jeffries2014}
{Jeffries} R.~D.,  2014, in EAS Publications Series. pp 289--325 (\mn@eprint
  {arXiv} {1404.7156}), \mn@doi{10.1051/eas/1465008}

\bibitem[\protect\citeauthoryear{{Jerabkova}, {Boffin}, {Beccari}  \&
  {Anderson}}{{Jerabkova} et~al.}{2019}]{Jerabkova2019}
{Jerabkova} T.,  {Boffin} H. M.~J.,  {Beccari} G.,   {Anderson} R.~I.,  2019,
  \mn@doi [\mnras] {10.1093/mnras/stz2315}, \href
  {https://ui.adsabs.harvard.edu/abs/2019MNRAS.489.4418J} {489, 4418}

\bibitem[\protect\citeauthoryear{{Jofr{\'e}} et~al.,}{{Jofr{\'e}}
  et~al.}{2014}]{Jofre2014}
{Jofr{\'e}} P.,  et~al., 2014, \mn@doi [\aap] {10.1051/0004-6361/201322440},
  \href {http://adsabs.harvard.edu/abs/2014A%26A...564A.133J} {564, A133, Paper
  III}

\bibitem[\protect\citeauthoryear{{Kharchenko}, {Piskunov}, {R{\"o}ser},
  {Schilbach}  \& {Scholz}}{{Kharchenko} et~al.}{2005}]{Kharchenko2005}
{Kharchenko} N.~V.,  {Piskunov} A.~E.,  {R{\"o}ser} S.,  {Schilbach} E.,
  {Scholz} R.~D.,  2005, \mn@doi [\aap] {10.1051/0004-6361:20052740}, \href
  {https://ui.adsabs.harvard.edu/abs/2005A&A...440..403K} {440, 403}

\bibitem[\protect\citeauthoryear{{Kobayashi} \& {Nakasato}}{{Kobayashi} \&
  {Nakasato}}{2011}]{Kobayashi2011}
{Kobayashi} C.,  {Nakasato} N.,  2011, \mn@doi [\apj]
  {10.1088/0004-637X/729/1/16}, \href
  {http://adsabs.harvard.edu/abs/2011ApJ...729...16K} {729, 16}

\bibitem[\protect\citeauthoryear{{Kobayashi}, {Umeda}, {Nomoto}, {Tominaga}  \&
  {Ohkubo}}{{Kobayashi} et~al.}{2006}]{Kobayashi2006}
{Kobayashi} C.,  {Umeda} H.,  {Nomoto} K.,  {Tominaga} N.,   {Ohkubo} T.,
  2006, \mn@doi [\apj] {10.1086/508914}, \href
  {http://adsabs.harvard.edu/abs/2006ApJ...653.1145K} {653, 1145}

\bibitem[\protect\citeauthoryear{{Kounkel} \& {Covey}}{{Kounkel} \&
  {Covey}}{2019}]{Kounkel2019}
{Kounkel} M.,  {Covey} K.,  2019, \mn@doi [\aj] {10.3847/1538-3881/ab339a},
  \href {https://ui.adsabs.harvard.edu/abs/2019AJ....158..122K} {158, 122}

\bibitem[\protect\citeauthoryear{{Kraus}, {Shkolnik}, {Allers}  \&
  {Liu}}{{Kraus} et~al.}{2014}]{Kraus2004}
{Kraus} A.~L.,  {Shkolnik} E.~L.,  {Allers} K.~N.,   {Liu} M.~C.,  2014,
  \mn@doi [\aj] {10.1088/0004-6256/147/6/146}, \href
  {https://ui.adsabs.harvard.edu/abs/2014AJ....147..146K} {147, 146}

\bibitem[\protect\citeauthoryear{{Krumholz} \& {Ting}}{{Krumholz} \&
  {Ting}}{2018}]{Krumholz2018}
{Krumholz} M.~R.,  {Ting} Y.-S.,  2018, \mn@doi [\mnras]
  {10.1093/mnras/stx3286}, \href
  {https://ui.adsabs.harvard.edu/abs/2018MNRAS.475.2236K} {475, 2236}

\bibitem[\protect\citeauthoryear{{Lada} \& {Lada}}{{Lada} \&
  {Lada}}{2003}]{Lada2003}
{Lada} C.~J.,  {Lada} E.~A.,  2003, \mn@doi [\araa]
  {10.1146/annurev.astro.41.011802.094844}, \href
  {https://ui.adsabs.harvard.edu/abs/2003ARA&A..41...57L} {41, 57}

\bibitem[\protect\citeauthoryear{{Lee} et~al.,}{{Lee}
  et~al.}{2008}]{2008AJ....136.2022L}
{Lee} Y.~S.,  et~al., 2008, \mn@doi [\aj] {10.1088/0004-6256/136/5/2022}, \href
  {https://ui.adsabs.harvard.edu/abs/2008AJ....136.2022L} {136, 2022}

\bibitem[\protect\citeauthoryear{{Liu}, {Yong}, {Asplund}, {Ram{\'\i}rez}  \&
  {Mel{\'e}ndez}}{{Liu} et~al.}{2016}]{Liu2016}
{Liu} F.,  {Yong} D.,  {Asplund} M.,  {Ram{\'\i}rez} I.,   {Mel{\'e}ndez} J.,
  2016, \mn@doi [\mnras] {10.1093/mnras/stw247}, \href
  {https://ui.adsabs.harvard.edu/abs/2016MNRAS.457.3934L} {457, 3934}

\bibitem[\protect\citeauthoryear{{Luo} et~al.,}{{Luo} et~al.}{2015}]{Luo2015}
{Luo} A.-L.,  et~al., 2015, \mn@doi [Research in Astronomy and Astrophysics]
  {10.1088/1674-4527/15/8/002}, \href
  {http://adsabs.harvard.edu/abs/2015RAA....15.1095L} {15, 1095}

\bibitem[\protect\citeauthoryear{{Majewski} et~al.,}{{Majewski}
  et~al.}{2017}]{Majewski2017}
{Majewski} S.~R.,  et~al., 2017, \mn@doi [\aj] {10.3847/1538-3881/aa784d},
  \href {http://adsabs.harvard.edu/abs/2017AJ....154...94M} {154, 94}

\bibitem[\protect\citeauthoryear{{Mamajek} \& {Hillenbrand}}{{Mamajek} \&
  {Hillenbrand}}{2008}]{mamajek2008}
{Mamajek} E.~E.,  {Hillenbrand} L.~A.,  2008, \mn@doi [\apj] {10.1086/591785},
  \href {http://adsabs.harvard.edu/abs/2008ApJ...687.1264M} {687, 1264}

\bibitem[\protect\citeauthoryear{{Mart{\'\i}n}, {Lodieu}, {Pavlenko}  \&
  {B{\'e}jar}}{{Mart{\'\i}n} et~al.}{2018}]{Martin2018}
{Mart{\'\i}n} E.~L.,  {Lodieu} N.,  {Pavlenko} Y.,   {B{\'e}jar} V. J.~S.,
  2018, \mn@doi [\apj] {10.3847/1538-4357/aaaeb8}, \href
  {https://ui.adsabs.harvard.edu/abs/2018ApJ...856...40M} {856, 40}

\bibitem[\protect\citeauthoryear{{Masseron} et~al.,}{{Masseron}
  et~al.}{2014}]{Masseron2014}
{Masseron} T.,  et~al., 2014, \mn@doi [\aap] {10.1051/0004-6361/201423956},
  \href {http://adsabs.harvard.edu/abs/2014A%26A...571A..47M} {571, A47}

\bibitem[\protect\citeauthoryear{{Masseron}, {Merle}  \& {Hawkins}}{{Masseron}
  et~al.}{2016}]{Masseron2016}
{Masseron} T.,  {Merle} T.,   {Hawkins} K.,  2016, {BACCHUS: Brussels Automatic
  Code for Characterizing High accUracy Spectra}, Astrophysics Source Code
  Library (\mn@eprint {ascl} {1605.004}), \mn@doi{10.20356/C4TG6R}

\bibitem[\protect\citeauthoryear{{Meingast}, {Alves}  \&
  {F{\"u}rnkranz}}{{Meingast} et~al.}{2019}]{Meingast2019b}
{Meingast} S.,  {Alves} J.,   {F{\"u}rnkranz} V.,  2019, \mn@doi [\aap]
  {10.1051/0004-6361/201834950}, \href
  {https://ui.adsabs.harvard.edu/abs/2019A&A...622L..13M} {622, L13}

\bibitem[\protect\citeauthoryear{{Michaud}}{{Michaud}}{1986}]{Michaud1986}
{Michaud} G.,  1986, \mn@doi [\apj] {10.1086/164025}, \href
  {https://ui.adsabs.harvard.edu/abs/1986ApJ...302..650M} {302, 650}

\bibitem[\protect\citeauthoryear{{Mikolaitis}, {de Laverny}, {Recio-Blanco},
  {Hill}, {Worley}  \& {de Pascale}}{{Mikolaitis}
  et~al.}{2017}]{Mikolaitis2017}
{Mikolaitis} {\v{S}}.,  {de Laverny} P.,  {Recio-Blanco} A.,  {Hill} V.,
  {Worley} C.~C.,   {de Pascale} M.,  2017, \mn@doi [\aap]
  {10.1051/0004-6361/201629629}, \href
  {https://ui.adsabs.harvard.edu/abs/2017A&A...600A..22M} {600, A22}

\bibitem[\protect\citeauthoryear{{Mishenina}, {Kovtyukh}, {Soubiran},
  {Travaglio}  \& {Busso}}{{Mishenina} et~al.}{2002}]{Mishenina2002}
{Mishenina} T.~V.,  {Kovtyukh} V.~V.,  {Soubiran} C.,  {Travaglio} C.,
  {Busso} M.,  2002, \mn@doi [\aap] {10.1051/0004-6361:20021399}, \href
  {http://adsabs.harvard.edu/abs/2002A%26A...396..189M} {396, 189}

\bibitem[\protect\citeauthoryear{{Ness} et~al.,}{{Ness}
  et~al.}{2018}]{Ness2018}
{Ness} M.,  et~al., 2018, \mn@doi [\apj] {10.3847/1538-4357/aa9d8e}, \href
  {https://ui.adsabs.harvard.edu/abs/2018ApJ...853..198N} {853, 198}

\bibitem[\protect\citeauthoryear{{Nomoto}, {Kobayashi}  \& {Tominaga}}{{Nomoto}
  et~al.}{2013}]{Nomoto2013}
{Nomoto} K.,  {Kobayashi} C.,   {Tominaga} N.,  2013, \mn@doi [\araa]
  {10.1146/annurev-astro-082812-140956}, \href
  {http://adsabs.harvard.edu/abs/2013ARA%26A..51..457N} {51, 457}

\bibitem[\protect\citeauthoryear{{Pinsonneault}}{{Pinsonneault}}{2010}]{Pinsonneault2010}
{Pinsonneault} M.~H.,  2010, in {Charbonnel} C.,  {Tosi} M.,  {Primas} F.,
  {Chiappini} C.,  eds,  IAU Symposium Vol. 268, Light Elements in the
  Universe. pp 375--380 (\mn@eprint {arXiv} {1001.3864}),
  \mn@doi{10.1017/S1743921310004497}

\bibitem[\protect\citeauthoryear{{Plez}}{{Plez}}{2012}]{Plez2012}
{Plez} B.,  2012, {Turbospectrum: Code for spectral synthesis}, Astrophysics
  Source Code Library (\mn@eprint {ascl} {1205.004})

\bibitem[\protect\citeauthoryear{{Ram{\'{\i}}rez}, {Mel{\'e}ndez}  \&
  {Chanam{\'e}}}{{Ram{\'{\i}}rez} et~al.}{2012}]{Ramirez2012}
{Ram{\'{\i}}rez} I.,  {Mel{\'e}ndez} J.,   {Chanam{\'e}} J.,  2012, \mn@doi
  [\apj] {10.1088/0004-637X/757/2/164}, \href
  {http://adsabs.harvard.edu/abs/2012ApJ...757..164R} {757, 164}

\bibitem[\protect\citeauthoryear{{Ratzenb{\"o}ck}, {Meingast}, {Alves},
  {M{\"o}ller}  \& {Bomze}}{{Ratzenb{\"o}ck} et~al.}{2020}]{Ratzenbock2020}
{Ratzenb{\"o}ck} S.,  {Meingast} S.,  {Alves} J.,  {M{\"o}ller} T.,   {Bomze}
  I.,  2020, arXiv e-prints, \href
  {https://ui.adsabs.harvard.edu/abs/2020arXiv200205728R} {p. arXiv:2002.05728}

\bibitem[\protect\citeauthoryear{{R{\"o}ser} \& {Schilbach}}{{R{\"o}ser} \&
  {Schilbach}}{2020}]{Roser2020}
{R{\"o}ser} S.,  {Schilbach} E.,  2020, arXiv e-prints, \href
  {https://ui.adsabs.harvard.edu/abs/2020arXiv200203610R} {p. arXiv:2002.03610}

\bibitem[\protect\citeauthoryear{{Samland}}{{Samland}}{1998}]{Samland1998}
{Samland} M.,  1998, \mn@doi [\apj] {10.1086/305368}, \href
  {http://adsabs.harvard.edu/abs/1998ApJ...496..155S} {496, 155}

\bibitem[\protect\citeauthoryear{{Sestito}, {Randich}, {Mermilliod}  \&
  {Pallavicini}}{{Sestito} et~al.}{2003}]{Sestito2003}
{Sestito} P.,  {Randich} S.,  {Mermilliod} J.~C.,   {Pallavicini} R.,  2003,
  \mn@doi [\aap] {10.1051/0004-6361:20030723}, \href
  {https://ui.adsabs.harvard.edu/abs/2003A&A...407..289S} {407, 289}

\bibitem[\protect\citeauthoryear{{Skumanich}}{{Skumanich}}{1972}]{Skumanich1972}
{Skumanich} A.,  1972, \mn@doi [\apj] {10.1086/151310}, \href
  {https://ui.adsabs.harvard.edu/abs/1972ApJ...171..565S} {171, 565}

\bibitem[\protect\citeauthoryear{{Sneden}}{{Sneden}}{1973}]{Sneden1973}
{Sneden} C.~A.,  1973, PhD thesis, THE UNIVERSITY OF TEXAS AT AUSTIN.

\bibitem[\protect\citeauthoryear{{Soderblom}}{{Soderblom}}{2010}]{Soderblom2010}
{Soderblom} D.~R.,  2010, \mn@doi [\araa]
  {10.1146/annurev-astro-081309-130806}, \href
  {http://adsabs.harvard.edu/abs/2010ARA%26A..48..581S} {48, 581}

\bibitem[\protect\citeauthoryear{{Soderblom}, {Laskar}, {Valenti}, {Stauffer}
  \& {Rebull}}{{Soderblom} et~al.}{2009}]{Soderblom2009}
{Soderblom} D.~R.,  {Laskar} T.,  {Valenti} J.~A.,  {Stauffer} J.~R.,
  {Rebull} L.~M.,  2009, \mn@doi [\aj] {10.1088/0004-6256/138/5/1292}, \href
  {https://ui.adsabs.harvard.edu/abs/2009AJ....138.1292S} {138, 1292}

\bibitem[\protect\citeauthoryear{{Takeda}, {Honda}, {Ohnishi}, {Ohkubo},
  {Hirata}  \& {Sadakane}}{{Takeda} et~al.}{2013}]{Takeda2013}
{Takeda} Y.,  {Honda} S.,  {Ohnishi} T.,  {Ohkubo} M.,  {Hirata} R.,
  {Sadakane} K.,  2013, \mn@doi [\pasj] {10.1093/pasj/65.3.53}, \href
  {https://ui.adsabs.harvard.edu/abs/2013PASJ...65...53T} {65, 53}

\bibitem[\protect\citeauthoryear{{Tull}, {MacQueen}, {Sneden}  \&
  {Lambert}}{{Tull} et~al.}{1995}]{Tull1995}
{Tull} R.~G.,  {MacQueen} P.~J.,  {Sneden} C.,   {Lambert} D.~L.,  1995,
  \mn@doi [\pasp] {10.1086/133548}, \href
  {https://ui.adsabs.harvard.edu/abs/1995PASP..107..251T} {107, 251}

\bibitem[\protect\citeauthoryear{{Tutukov}}{{Tutukov}}{1978}]{Tutukov1978}
{Tutukov} A.~V.,  1978, \aap, \href
  {https://ui.adsabs.harvard.edu/abs/1978A&A....70...57T} {70, 57}

\bibitem[\protect\citeauthoryear{{Valenti} \& {Fischer}}{{Valenti} \&
  {Fischer}}{2005}]{Valenti2005}
{Valenti} J.~A.,  {Fischer} D.~A.,  2005, \mn@doi [\apjs] {10.1086/430500},
  \href {http://adsabs.harvard.edu/abs/2005ApJS..159..141V} {159, 141}

\bibitem[\protect\citeauthoryear{{Valenti} \& {Piskunov}}{{Valenti} \&
  {Piskunov}}{1996}]{sme}
{Valenti} J.~A.,  {Piskunov} N.,  1996, \aaps, \href
  {http://adsabs.harvard.edu/abs/1996A\%26AS..118..595V} {118, 595}

\bibitem[\protect\citeauthoryear{{Xiang} et~al.,}{{Xiang}
  et~al.}{2017}]{Xiang2017}
{Xiang} M.-S.,  et~al., 2017, \mn@doi [\mnras] {10.1093/mnras/stx129}, \href
  {http://adsabs.harvard.edu/abs/2017MNRAS.467.1890X} {467, 1890}

\bibitem[\protect\citeauthoryear{{Xiang} et~al.,}{{Xiang}
  et~al.}{2019}]{Xiang2019}
{Xiang} M.,  et~al., 2019, \mn@doi [\apjs] {10.3847/1538-4365/ab5364}, \href
  {https://ui.adsabs.harvard.edu/abs/2019ApJS..245...34X} {245, 34}

\bibitem[\protect\citeauthoryear{{Yana Galarza} et~al.,}{{Yana Galarza}
  et~al.}{2019}]{JYG2019}
{Yana Galarza} J.,  et~al., 2019, \mn@doi [\mnras] {10.1093/mnrasl/slz153},
  \href {https://ui.adsabs.harvard.edu/abs/2019MNRAS.490L..86Y} {490, L86}

\bibitem[\protect\citeauthoryear{{van de Voort}, {Pakmor}, {Grand },
  {Springel}, {G{\'o}mez}  \& {Marinacci}}{{van de Voort}
  et~al.}{2019}]{Vandevoort2019}
{van de Voort} F.,  {Pakmor} R.,  {Grand } R. J.~J.,  {Springel} V.,
  {G{\'o}mez} F.~A.,   {Marinacci} F.,  2019, arXiv e-prints, \href
  {https://ui.adsabs.harvard.edu/abs/2019arXiv190701557V} {p. arXiv:1907.01557}

\makeatother
\end{thebibliography}


\bsp	
\label{lastpage}
\end{document}